\documentclass[camera,letterpaper,nomarginnotes,nonarrowgutter]{jpaper}

\usepackage{cite}
\usepackage{amsmath,amssymb,amsfonts}
\usepackage{textcomp}
\usepackage{tabularx}
\usepackage{array}
\usepackage{xurl}
\usepackage{tikz}
\usepackage{soul}
\usepackage{tcolorbox}
\usepackage{balance}
\usepackage{fancyhdr}
\usepackage{hyperref}
\usepackage{dblfloatfix}
\newcommand{\boxedurl}[1]{%
  \begingroup
  \setlength{\fboxsep}{1pt}%
  \fcolorbox{cyan}{white}{\href{#1}{\nolinkurl{#1}}}%
  \endgroup
}
\hypersetup{hidelinks}

\sethlcolor{gray!20}

\newcommand{\cmd}[1]{%
  \begingroup
  \ttfamily
  \hl{#1}%
  \endgroup
}

\newcommand{\artifactbadges}{%
  \begin{tikzpicture}[remember picture,overlay]
    \node[
      anchor=north east,
      xshift=-1.0cm,
      yshift=-0.15cm
    ] at (current page.north east) {%
      \includegraphics[height=1.5cm]{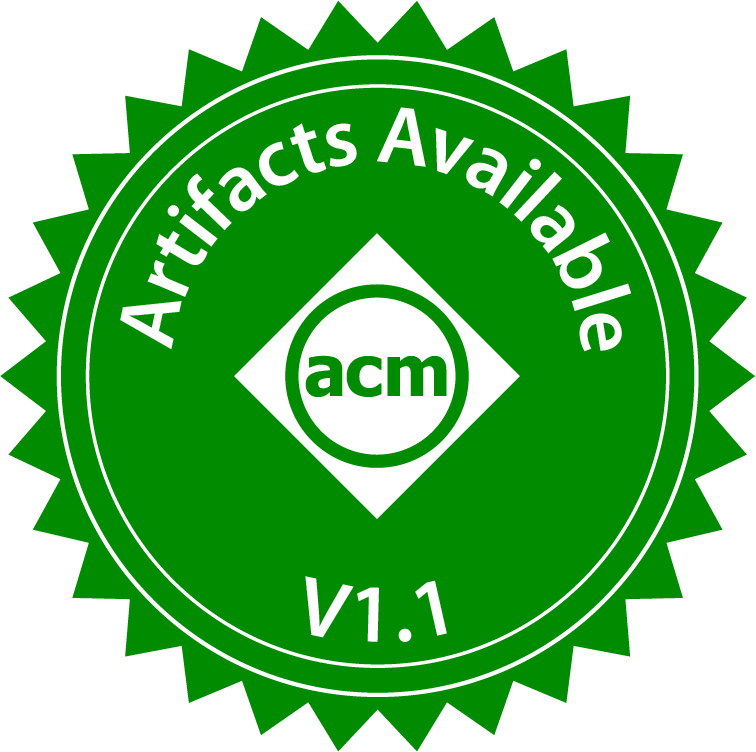}%
      \hspace{2pt}%
      \includegraphics[height=1.5cm]{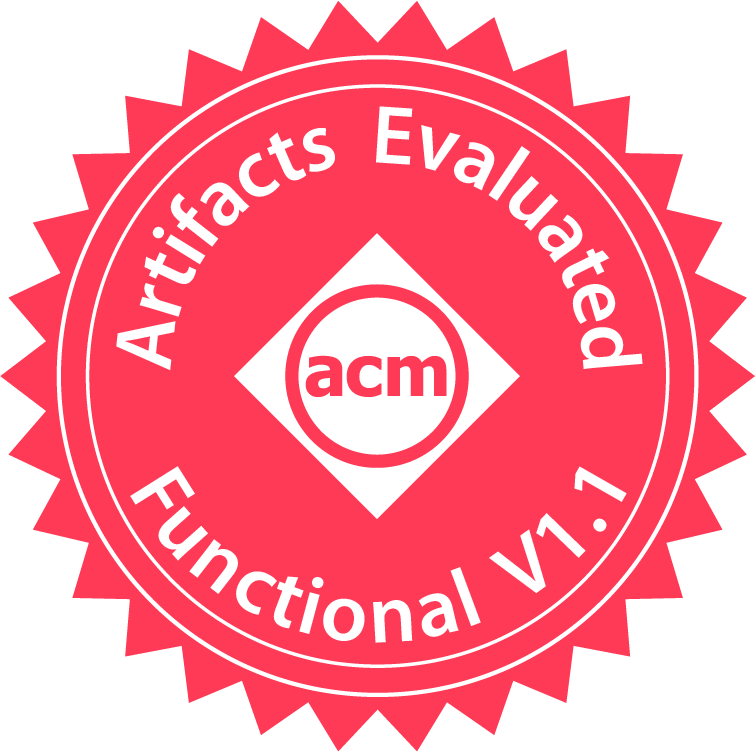}%
      \hspace{2pt}%
      \includegraphics[height=1.5cm]{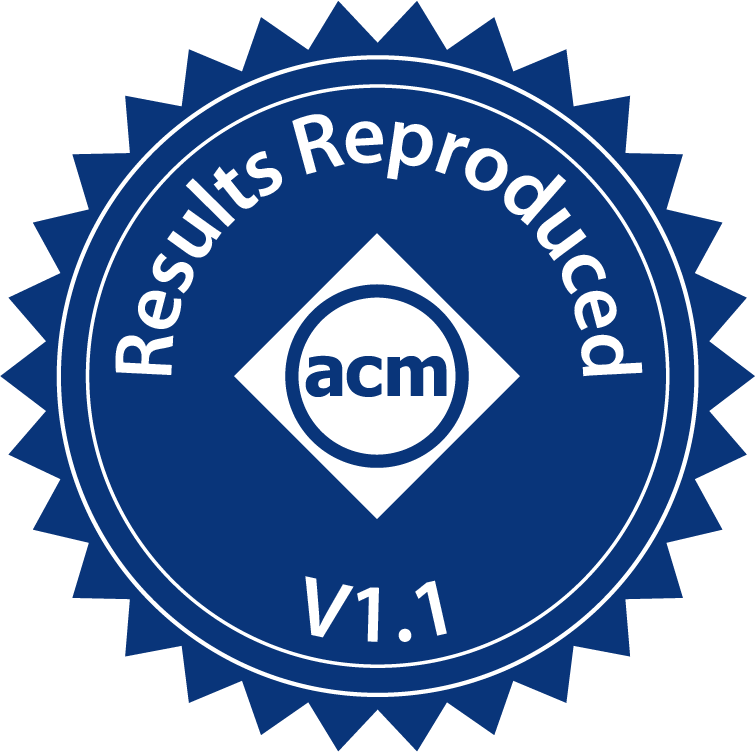}%
    };
  \end{tikzpicture}%
}

\begin{document}
\bstctlcite{IEEEexample:BSTcontrol}
\title{%
  \fontsize{18}{19}\selectfont
  MORDOR: \underline{M}itigating \underline{O}verheads
  of \underline{R}ead \underline{D}isturbance Preventive \underline{O}perations
  via Elastic \underline{R}efresh Scheduling%
}
\author{%
  \large
  \begin{tabular}{c@{\hspace{1.8em}}c@{\hspace{1.8em}}c}
    Maria Makeenkova\textsuperscript{\S} &
    Ataberk Olgun\textsuperscript{\S} &
    F. Nisa Bostanc{\i}\textsuperscript{\S} \\[0.25ex]
    İsmail Emir Y\"{u}ksel\textsuperscript{\S} &
    Spiros Galanopoulos\textsuperscript{\S} &
    Onur Mutlu\textsuperscript{\S\textdagger}
  \end{tabular}
  \\[1.5ex]
  \textsuperscript{\S}\textit{ETH Zurich}
  \qquad
  \textsuperscript{\textdagger}\textit{New York University}
}

\newcommand{\X}[0]{MORDOR}

\newcommand{\xp}[0]{$P_{adv}$}

\newcommand{\nrh}[0]{$N_{\mathrm{RH}}$}

\newcommand{\circled}[1]{\tikz[baseline=(char.base)]{
    \node[shape=circle,draw,inner sep=1pt] (char) {#1};}}

\newcommand{\ballnumber}[1]{\tikz[baseline=(myanchor.base)] \node[circle,fill=.,inner sep=1pt] (myanchor) {\color{-.}\bfseries\footnotesize #1};}

\newcommand{\ramulatortwocite}[0]{\cite{luo2023ramulator, ramulator2github, bostanci2026cleaning, kim2016ramulator}}

\newcommand{\mitigatingRowHammerAllCitations}[0]{\cite{AppleRefInc, rh-hp, rh-lenovo, greenfield2012throttling, kim2014flipping, kim2015architectural, bains14d, bains14c, bains14a,
  bains14b, aweke2016anvil, bains2015row, bains2016row, son2017making, seyedzadeh2018cbt, irazoqui2016mascat, you2019mrloc, lee2019twice, park2020graphene, yaglikci2021security,
  yaglikci2021blockhammer, frigo2020trrespass, kang2020cattwo, hassan2021utrr, qureshi2022hydra, saileshwar2022randomized, brasser2017can, konoth2018zebram, van2018guardion, vig2018rapid,
  kim2022mithril, lee2021cryoguard, marazzi2022protrr, zhang2022softtrr, joardar2022learning, juffinger2023csi, yaglikci2022hira, saxena2022aqua, enomoto2022efficient,
  manzhosov2022revisiting, ajorpaz2022evax, naseredini2022alarm, joardar2022machine, hassan2022case, zhang2020leveraging, loughlin2021stop, devaux2021method, fakhrzadehgan2022safeguard,
  saroiu2022price, loughlin2022moesiprime, han2021surround, woo2022scalable, bock2019riprh, wang2021discreet, bennett2021panopticon, kim2023ddr5, qureshi2024mint, kwong2020rambleed,
  cojocar2020rowhammer, weissman2020jackhammer, zhang2020pthammer, yao2020deephammer, deridder2021smash, jattke2022blacksmith, tol2022toward, kogler2022half, orosa2022spyhammer,
  zhang2022implicit, liu2022generating, cohen2022hammerscope, zheng2022trojvit, fahr2022frodo, tobah2022spechammer, rakin2022deepsteal, hajiabadi2025charm, yauglikcci2021security,
  bostanci2024comet, olgun2024abacus, dapper, saxena2024start, seyedzadeh2017cbt, Nale2019, lin2015handling, taneja2025dream, kim2021hammerfilter, saroiu2022configure, vittal2025mopac,
  stefanDRFM, yaglikci2024svard, saxena2024impress, hong2023dsac,
  jaleel2024pride, moat, qprac, canpolat2025chronus, woo2025mitigations, qureshi2025autorfm, taneja2026mirza, wu2026apt, lin2025cnc}}

\newcommand{\mitigatingRowHammerCounterCitations}[0]{\cite{kim2014flipping, seyedzadeh2018cbt, park2020graphene, kim2022mithril, marazzi2022protrr, yaglikci2022hira, loughlin2021stop,
  bennett2021panopticon, kim2023ddr5, qureshi2022hydra, greenfield2012throttling, kim2015architectural, bains14d, bains14c, bains14a, bains14b, aweke2016anvil, bains2015row, bains2016row,
  son2017making, you2019mrloc, lee2019twice, yaglikci2021security, kang2020cattwo, joardar2022learning, devaux2021method, wang2021discreet, bostanci2024comet, olgun2024abacus, dapper,
  saxena2024start, seyedzadeh2017cbt, hajiabadi2025charm, Nale2019, lin2015handling, taneja2025dream, kim2021hammerfilter, saroiu2022configure, vittal2025mopac, stefanDRFM,
  yaglikci2024svard, saxena2024impress, qureshi2024mint, hong2023dsac,
  jaleel2024pride, moat, qprac, canpolat2025chronus, woo2025mitigations, qureshi2025autorfm, taneja2026mirza, wu2026apt, lin2025cnc}}

\newcommand{\mitigatingRowHammerMCAllCitations}[0]{\cite{kim2014flipping, seyedzadeh2018cbt, park2020graphene, yaglikci2022hira, loughlin2021stop, qureshi2022hydra,
  greenfield2012throttling, kim2015architectural, bains14d, bains14c, bains14a, bains14b, bains2015row, bains2016row, son2017making, you2019mrloc, yaglikci2021security, kang2020cattwo,
  joardar2022learning, wang2021discreet, bostanci2024comet, olgun2024abacus, dapper, saxena2024start, seyedzadeh2017cbt, hajiabadi2025charm, Nale2019, lin2015handling, taneja2025dream,
  kim2021hammerfilter, saroiu2022configure, vittal2025mopac, stefanDRFM, yaglikci2024svard, saxena2024impress}}

\newcommand{\mitigatingRowHammerMCCounterCitations}[0]{\cite{seyedzadeh2018cbt, park2020graphene, qureshi2022hydra, greenfield2012throttling, kim2015architectural, bains14d, bains14c,
  bains14a, bains14b, bains2015row, bains2016row, kang2020cattwo, joardar2022learning, bostanci2024comet, olgun2024abacus, dapper, saxena2024start, seyedzadeh2017cbt, hajiabadi2025charm,
  Nale2019, lin2015handling, taneja2025dream, kim2021hammerfilter, stefanDRFM, yaglikci2024svard, saxena2024impress}}

\newcommand{\probabilisticRowHammerDefenseCitations}[0]{\cite{kim2014flipping, you2019mrloc, son2017making, wang2021discreet, kim2015architectural, saroiu2022configure, vittal2025mopac,
  kim2021hammerfilter}}

\newcommand{\exploitingRowHammerNewCitations}[0]{\cite{kim2014flipping,mutlu2019rowhammer,kwong2020rambleed, frigo2020trrespass, cojocar2020rowhammer, weissman2020jackhammer, zhang2020pthammer, yao2020deephammer, deridder2021smash, hassan2021utrr, jattke2022blacksmith, tol2022toward, kogler2022half, orosa2022spyhammer, zhang2022implicit, liu2022generating, cohen2022hammerscope, zheng2022trojvit, fahr2022frodo, tobah2022spechammer, rakin2022deepsteal,bhattacharya2018advanced,google-project-zero,razavi2016flip,pessl2016drama,xiao2016one,bosman2016dedup,bhattacharya2016curious,burleson2016invited,qiao2016new,brasser2017can,jang2017sgx,aga2017good,tatar2018defeating,gruss2018another,mutlu2017rowhammer,lipp2018nethammer,van2018guardion,frigo2018grand,cojocar2019eccploit,ji2019pinpoint,mutlu2019rowhammerandbeyond,hong2019terminal,fournaris2017exploiting,poddebniak2018attacking,tatar2018throwhammer,carre2018openssl,barenghi2018software,zhang2018triggering,rowhammergithub,kang2024sledgehammer,lin2025gpuhammer,nazaraliyev2025notsorefreshing,meyer2026phoenix,kamadan2025ecc}}

\newcommand{\proimmediately}[0]{\cite{yaglikci2022hira, bennett2021panopticon, qureshi2022hydra}}

\newcommand{\secref}[1]{§\ref{#1}}

\maketitle
\artifactbadges
\begin{abstract}
Modern DRAM chips are susceptible to read disturbance phenomena such as RowHammer, where repeatedly accessing (hammering) a row of DRAM cells (i.e., a DRAM row) induces bitflips in other physically nearby (victim) DRAM rows.
A common practice to avoid such bitflips is to \emph{preventively refresh} victim rows that might otherwise experience 
bitflips.
Unfortunately, preventive refreshes cause long latencies
and need to be performed urgently before the aggressor row is activated again to ensure data integrity. This is done by prioritizing them over demand memory requests, thereby potentially imposing significant delays on those requests and causing performance and energy overheads. \leavevmode{Our goal in this work is to alleviate these overheads by scheduling preventive refreshes off the critical path of demand memory requests.}
We propose \X{}, a new preventive refresh scheduling policy that
significantly reduces system performance degradation and energy consumption caused by preventive refresh operations.
\leavevmode{\X{} is integrated into the memory controller and operates alongside memory-controller-based read disturbance mitigation techniques to intelligently delay preventive refresh operations, 
while maintaining
their data integrity guarantees.} \X{} leverages the key observation that a preventive refresh operation
targeting an aggressor row 
\leavevmode{can be delayed to serve any other demand memory request, as long as that memory request does \emph{not} access the aggressor row.}
By doing so, \X{} executes latency-critical
memory requests before long-latency preventive refresh operations, while mitigating read disturbance bitflips. We 
evaluate \X{}
by integrating it into six state-of-the-art read disturbance mitigation techniques.
Our comprehensive evaluation shows
that \X{} significantly improves system performance and energy efficiency
at low area cost,  by reducing the latencies of demand memory requests, \textcolor{black}{while maintaining the data integrity guarantees of the underlying read disturbance mitigation technique}.
Across 55 eight-core workloads, \X{} provides 
an average (maximum) 21.7\% ({423.0}\%) speedup and a {22.1}\% (\leavevmode{411.5}\%) 
DRAM energy reduction when 
integrated with {four} state-of-the-art read disturbance mitigation techniques {that issue a significant number of preventive refresh operations}
for a preventive refresh threshold of 125. \X{} is openly and freely available at \boxedurl{https://github.com/CMU-SAFARI/MORDOR}.

\end{abstract}

\section{Introduction}
Modern DRAM chips are vulnerable to read disturbance \cite{kim2014flipping, mutlu2017rowhammer, yang2019trap, mutlu2019rowhammer,park2016statistical, park2016experiments,
walker2021ondramrowhammer, ryu2017overcoming, yang2016suppression, yang2017scanning, gautam2019row, jiang2021quantifying,mutlu2023fundamentally, kim2020revisiting, olgun2023hbm, luo2023rowpress}, where repeatedly activating and precharging (opening and closing) an aggressor DRAM row (hammering) at a sufficiently high rate can cause bitflips in its physically adjacent DRAM rows (called victim rows). This vulnerability is known as RowHammer and poses a serious threat to the robustness and security of DRAM-based computing systems. Prior works have shown that RowHammer can be maliciously exploited for privilege escalation, or to cause data leakage and corruption\leavevmode{~\exploitingRowHammerNewCitations}.
Consequently, various techniques have been proposed to prevent read-disturbance-induced bitflips\leavevmode{~\mitigatingRowHammerAllCitations{}}. 
\leavevmode{Many} of these works (e.g.,\leavevmode{~\mitigatingRowHammerCounterCitations{}}) propose using Preventive Refresh Operations (PROs). A PRO is issued to a row that is identified as an aggressor row by the read disturbance mitigation technique.
\leavevmode{Recent} works~\cite{kim2014flipping, mutlu2017rowhammer, mutlu2019retrospective, kim2020revisiting, yaglikci2022understanding, orosa2021deeper, mutlu2018rowhammer, mutlu2023fundamentally} show that as the technology node size reduces, bitflips can occur with even fewer DRAM row activations. Specifically, \cite{kim2020revisiting} shows that the number of activations needed to induce bitflips in neighboring rows has reduced by more than an order of magnitude in less than a decade. As a result, Preventive Refresh Threshold (PRT) values have also reduced, causing PROs to happen more frequently. 

Prior works~\textcolor{black}{\mitigatingRowHammerMCAllCitations{}} propose a \leavevmode{naïve} PRO scheduling approach that mitigates read disturbance bitflips at the cost of higher memory access latency: they prioritize PROs over all other requests in the memory controller queues \textcolor{black}{\proimmediately{}}. \leavevmode{Doing so overrides the schedule of memory operations decided by the memory scheduling scheme, e.g., First-Ready First-Come First-Serve (FR-FCFS)~\cite{rixner2000memory, zuravleff1997controller}, PAR-BS~\cite{mutlu2009parallelism}, or STFM~\cite{mutlu2007stall}, \leavevmode{and slows down all other memory requests that are waiting to be scheduled by the memory controller.}} As PRT values decrease, mitigation techniques incur higher execution time and energy consumption overheads due \leavevmode{to a significant increase in the number of required preventive refreshes.}

\textbf{Our goal} in this work is to improve performance and reduce energy consumption of memory-controller-based read disturbance mitigation techniques, especially at low PRT values. To this end, we propose \X{}, a new PRO scheduling algorithm that leverages the following \leavevmode{\textbf{key observation}}:
\begin{tcolorbox}[
  boxsep=3pt,
  left=3pt,
  right=3pt,
  top=3pt,
  bottom=3pt,
  before skip=6pt,
  after skip=6pt 
]
A PRO that is issued to an aggressor DRAM row must \textit{only} be prioritized over another request if that request activates the \leavevmode{\emph{same}} aggressor row.
\end{tcolorbox}

\leavevmode{\textbf{Key Idea. }\X{} exploits this observation by allowing requests that target \leavevmode{\emph{other}} rows \leavevmode{(than those requiring PROs)} to continue to be scheduled according to the memory controller's underlying scheduling algorithm while a PRO is pending.}

\leavevmode{\textbf{Mechanism. }To prevent an aggressor row from being reactivated before its PRO completes, \X{} temporarily blacklists rows with outstanding PROs. As a result, demand requests that do \leavevmode{\emph{not}} access a blacklisted row can continue to make progress, reducing interference caused by PROs and thereby reducing the average memory access latency, execution time, and energy consumption.}

\leavevmode{\textbf{Implementation.}} \X{} is integrated into a memory controller design with two minor modifications. \leavevmode{(1)}~The Preventive Refresh Operation Queue (PROQ), which serves as a blacklist and stores in-flight PROs. \X{} queries the PROQ to determine if issuing a memory request could cause a read disturbance bitflip.
If the request targets a blacklisted DRAM row (i.e., one that can be found in the PROQ), the request cannot proceed. 
To perform such queries quickly, we implement the PROQ as a small (120 bytes of storage) content-addressable memory. \leavevmode{(2)~}A 50-byte Request Age Counter (RAC) indicates the relative age of demand memory requests and PROs enqueued in the memory controller. \leavevmode{This ordering is used by the memory scheduler for scheduling decisions.}

\textbf{Key Results.} 
\leavevmode{We evaluate \X{}'s impact on system performance and energy consumption alongside six state-of-the-art read disturbance mitigation techniques ~\cite{kim2014flipping, qureshi2022hydra, bostanci2024comet, dapper, park2020graphene, olgun2024abacus} for a range of} \leavevmode{RowHammer Threshold (\nrh{}) values} \leavevmode{from 125 to 1000.}
\leavevmode{Our evaluation uses a cycle-accurate memory system simulator, Ramulator 2.0~\textcolor{black}{\ramulatortwocite{}}. We evaluate a diverse set of 55 8-core homogeneous workloads from SPEC CPU2006~\cite{spec2006}, SPEC CPU2017~\cite{spec2017}, {TPC~\cite{tpcweb}, MediaBench~\cite{fritts2009media}, and YCSB~\cite{ycsb}} benchmark suites, and memory-intensive microbenchmarks.}
Across all read disturbance mitigation techniques and \leavevmode{\nrh{} values}, we observe an average speedup of \leavevmode{8.6\%} and an average reduction of \leavevmode{8.1\%} in \leavevmode{DRAM} energy consumption when using \X{} compared to when PROs are prioritized. The greatest benefits of \X{} are observed at \leavevmode{an \nrh{} value} of 125 with an average speedup of \leavevmode{14.5\%} and a \leavevmode{14.8\%} reduction in energy consumption. This is achievable at a very small storage overhead of 170B.

\leavevmode{We evaluate the security guarantees of \X{} and demonstrate that \X{} fully preserves the data integrity guarantees of the underlying read disturbance mitigation techniques and does  \leavevmode{\emph{not}} introduce a new angle of attack by delaying PROs.}
\leavevmode{We further analyze \X{}'s behavior when the fixed sized PROQ reaches its capacity. We observe that a 32-entry PROQ is sufficient for the vast majority of evaluated workloads. At rare times when the PROQ fills up, including adversarial access patterns designed to induce large numbers of outstanding PROs, \X{} safely falls back to prioritizing PROs until capacity becomes available.}
\leavevmode{We open source our simulation infrastructure and datasets at \boxedurl{https://github.com/CMU-SAFARI/MORDOR} to enable reproducibility and help future research.}

In this work, we make the following \textbf{key contributions:}
\begin{itemize}
    \item We provide the insight that a \leavevmode{Preventive Refresh Operation (PRO)} does \leavevmode{\emph{not}} have to be prioritized over other memory requests, as long as those requests do \leavevmode{\emph{not}} open the target aggressor row of the PRO.
    \item We develop \X{}, a mechanism that leverages the above observation to schedule PROs off the critical path of memory requests that do \leavevmode{\emph{not}} open the PRO's target aggressor row. Doing so enables the memory controller to better follow the underlying scheduling algorithm when issuing memory requests, while still preventing read disturbance bitflips.
    \item We evaluate the performance, energy, and area overheads of \X{} implemented with six state-of-the-art read disturbance mitigation techniques at four different \leavevmode{\nrh{}} values. We show that \X{} significantly reduces the overheads of four of those read disturbance mitigation techniques, while maintaining security against read disturbance-induced bitflips.
    \item We show how \X{} can be implemented \leavevmode{with a very low hardware overhead \leavevmode{(e.g.\leavevmode{,} 170B)} for both benign and adversarial memory access patterns}.
    \item  \leavevmode{Our security analysis shows that \X{} can be implemented alongside any memory-controller-based read disturbance mitigation technique that uses PROs to refresh potential victim rows without compromising its data integrity guarantees.}
\end{itemize}
\section{Background}
\begin{figure}[h]
    \centering
    \includegraphics[width=0.96\linewidth]{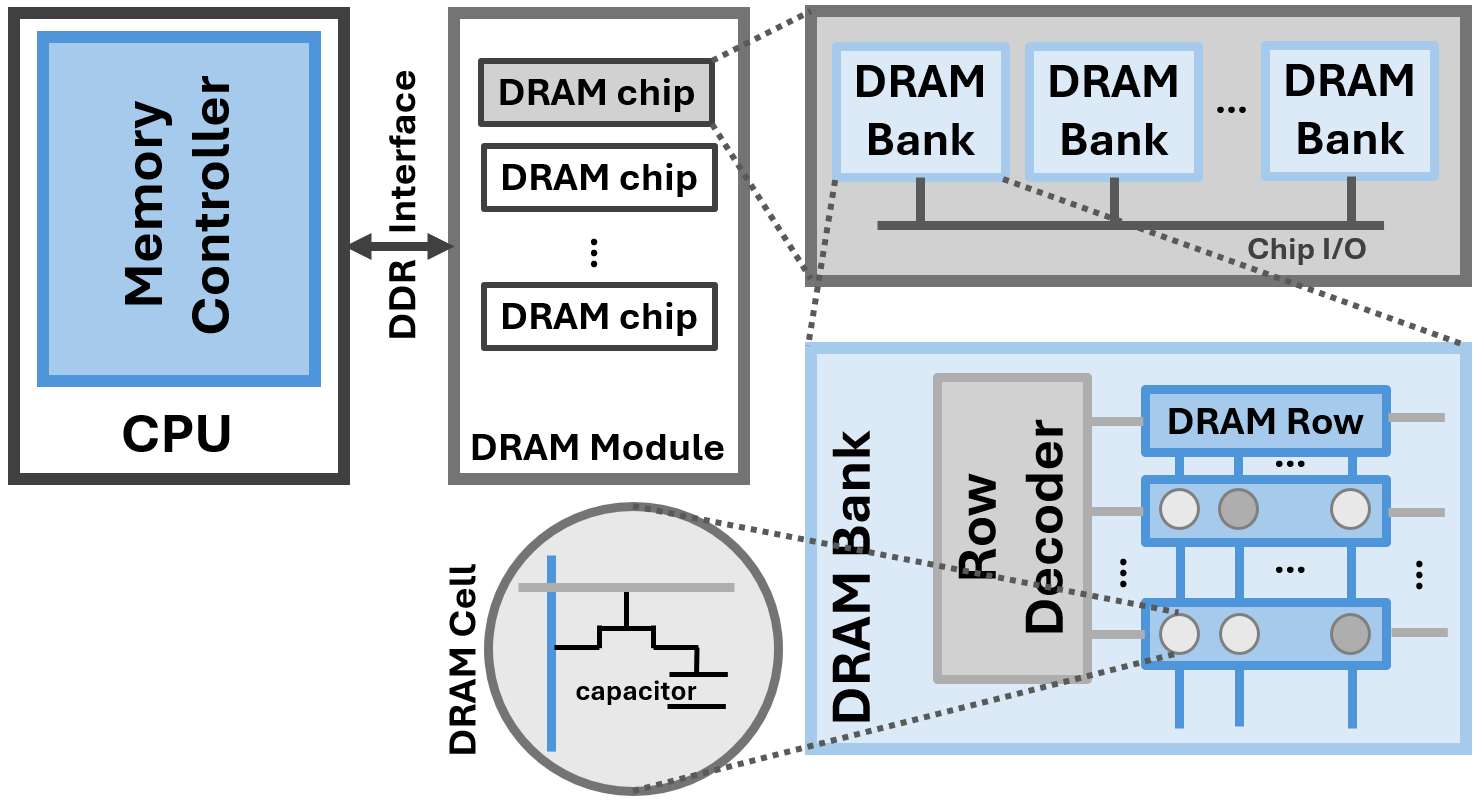}
    \caption{DRAM organization}
    \label{fig:dram_org}
\end{figure}

\textbf{DRAM Organization. }
Figure~\ref{fig:dram_org} shows the organization of a DRAM-based memory system. The CPU's memory controller is connected to the DRAM module over a DDR interface. The DRAM module consists of multiple DRAM chips, made up of DRAM banks that are connected by chip I/O. DRAM banks consist of DRAM rows, which are made up of interconnected DRAM cells. Each DRAM cell is a capacitor, whose charge stores a single bit of data. A row decoder in each DRAM bank selects specific DRAM rows within the bank for read or write operations.

\textbf{DRAM Operation. }
To read from and write to memory, the memory controller issues a sequence of commands to DRAM. First, the memory controller issues an activate (\textit{ACT}) command to the bank and row corresponding to the requested memory address. This command opens the row, allowing the memory controller to read or write the data stored in the row's cells. A read (\textit{RD}) or write (\textit{WR}) command can then be issued to the row. To access another row in the same bank, the currently open row must be closed by issuing a precharge (\textit{PRE}) command. 

\textbf{Periodic Refresh. }Due to the volatile nature of the capacitor that stores the data value of each DRAM cell, DRAM cells leak charge over time\leavevmode{\cite{liu2012raidr, liu2013experimental}}. In order to prevent data loss due to charge leakage, DRAM cells need to be refreshed periodically. Such refreshes happen in every \textit{refresh window}, which is typically 64ms for DDR4\cite{jedec2017ddr4} or 32ms for DDR5\cite{jedec2020ddr5}. To refresh a row, the memory controller sends a refresh (\textit{REF}) command, which restores the charge in the DRAM cells \cite{harris_harris}.

\textbf{Read Disturbance Errors. }
\leavevmode{DRAM is vulnerable to \emph{read disturbance phenomena} where accessing} a DRAM row can cause electrical disturbance in nearby rows \emph{without} accessing those rows directly~\cite{kim2014flipping}. 
\leavevmode{RowHammer is a prominent example of read disturbance, where repeatedly activating and precharging an aggressor row can induce bitflips in physically nearby victim rows. RowHammer-induced bitflips occur only after an aggressor row is activated sufficiently many times within a limited time interval.  We refer to the minimum number of activations required to induce a RowHammer bitflip as the RowHammer threshold ($N_{\mathrm{RH}}$).} Due to increased chip density in modern DRAM chips, \leavevmode{$N_{\mathrm{RH}}$} has reduced, making DRAM chips significantly more vulnerable to read disturbance errors\leavevmode{~\cite{kim2014flipping, mutlu2017rowhammer, mutlu2019retrospective, kim2020revisiting, yaglikci2022understanding, orosa2021deeper, mutlu2018rowhammer, mutlu2023fundamentally, yaglikci2021blockhammer, bostanci2024comet, saxena2023randomized}}. 

\textbf{Read Disturbance Mitigation Techniques. }
\label{sec:mech_background}
\leavevmode{Many}
works propose mitigation techniques to prevent read-disturbance-induced bitflips\leavevmode{\mitigatingRowHammerAllCitations}. \leavevmode{To do so, {these} techniques identify potential aggressor rows and take a mitigating action before any aggressor {row} reaches $N_{\mathrm{RH}}$ activations. {A set of mitigation} 
techniques~\cite{kim2014flipping, seyedzadeh2018cbt, park2020graphene, kim2022mithril, marazzi2022protrr, yaglikci2022hira, loughlin2021stop,bennett2021panopticon,saxena2022aqua,saileshwar2022randomized,kim2023ddr5, qureshi2022hydra, qureshi2024mint} preventively refresh 
potential victim rows of an aggressor {row to prevent bitflips}. We refer to such an operation as a Preventive Refresh Operation (PRO). A PRO targets the aggressor row, and preventively refreshes its victim rows.
A mitigation technique triggers a PRO according to a technique-specific Preventive Refresh Threshold (PRT), which is chosen sufficiently below $N_{\mathrm{RH}}$ to preserve the mitigation technique’s data integrity guarantees.}
For an issued PRO to be effective in preventing read disturbance
bitflips, the aggressor row must \leavevmode{\emph{not}} receive any further activations until \leavevmode{the mitigation \leavevmode{(i.e., refresh)} occurs}. 
\leavevmode{Thus, many mitigation techniques prioritize PROs over all other requests in the memory controller queues to avoid activating an aggressor row with an outstanding PRO~\textcolor{black}{\proimmediately{}}.}

\leavevmode{We analyze and evaluate six state-of-the-art PRO-based RowHammer mitigation techniques that represent a diverse set of aggressor tracking approaches: 1)~probabilistic tracking, 2)~exact activation counting, and 3)~approximate activation counting with various methods.\footnote{We describe and discuss other RowHammer mitigation techniques in §\ref{sec:related_works}.}}

\textbf{PARA}
\cite{kim2014flipping} \leavevmode{protects against RowHammer by performing probabilistic adjacent row activation. When a row is closed (i.e., when the memory controller issues a precharge command), PARA issues
a PRO to that row based on a probability threshold $p$ which refreshes its adjacent victim rows. }

\textbf{Hydra}
\cite{qureshi2022hydra} \leavevmode{implements a
group count table to track activations for a group of rows and
a row count table to track per-row activations. The row count table is stored in
the DRAM and cached in the memory controller. Hydra issues \leavevmode{a PRO} when a counter exceeds the PRT value.}

\textbf{Graphene} \cite{park2020graphene} \leavevmode{implements per-bank counters to track the possible aggressor rows using the Misra-Gries algorithm\cite{misra1982finding}.  When a row's counter value exceeds the PRT value, Graphene issues a PRO to that row to refresh its victims.}

\textbf{ABACuS}
\cite{olgun2024abacus} \leavevmode{uses counters shared between rows with the same row address across different banks to track row activation counts. ABACuS also tracks which banks contribute to the shared count to more accurately determine each row's activation count. When a row's activation count reaches the PRT, ABACuS issues a PRO to that row to refresh its victims.}

\textbf{CoMeT}
\cite{bostanci2024comet} \leavevmode{uses a Count-Min Sketch \cite{cormode2009count} approach to track row activation counts, and maps each DRAM row to multiple shared counters using a set of hash functions. When a row's estimated activation count reaches the PRT, CoMeT issues a PRO to that row to refresh its victims.}

\textbf{DAPPER}
\cite{dapper} \leavevmode{maps each DRAM row to two Row Group Counters (RGCs) using two hash functions. Once both RGCs reach the PRT, DAPPER identifies the rows shared by the two row groups as potential aggressor rows and issues PROs to those rows. }

\textbf{Directed Refresh Management \leavevmode{(DRFM)}.}
\leavevmode{The DDR5 specification~\cite{jedec2020ddr5} introduces a DRAM command called Directed Refresh Management (DRFM) to enable preventively refreshing victim rows of an arbitrary row. Using this command, a mitigation technique can issue PROs to any row upon determining that a mitigating action is needed to prevent bitflips. A DRFM command can target 1)~all banks (DRFMab)\leavevmode{,} such that it refreshes the victims of a row across all banks, or 2)~same bank across all bank groups (DRFMsb)\leavevmode{,} such that it refreshes the victims of a row in a bank across all bank groups.}

\section{Motivational Analysis}
Many state-of-the-art \leavevmode{memory-controller-based} read disturbance mitigation techniques incur non-negligible performance overheads on the system. We hypothesize that this is in part due to the fact that \leavevmode{once a read disturbance mitigation technique detects an aggressor row}, a \leavevmode{Preventive Refresh Operation (PRO)} must be issued \emph{immediately}\leavevmode{~\proimmediately{}}. \leavevmode{The memory controller guarantees this} by prioritizing \leavevmode{the PRO} over other demand memory requests, to make sure that the aggressor row does \leavevmode{\emph{not}} receive any further activations before the victims are preventively refreshed. 
This \emph{Priority Scheduling} model is an assumption many state-of-the-art read disturbance mitigation techniques \textcolor{black}{\cite{qureshi2022hydra, bostanci2024comet, olgun2024abacus, dapper, park2020graphene, kim2014flipping, qureshi2024mint, bennett2021panopticon}} make \leavevmode{(either explicitly or \textcolor{black}{implicitly})} to \leavevmode{guarantee} \leavevmode{data integrity.}
{\leavevmode{However, Priority Scheduling} schedules PROs on the critical path of unrelated memory requests, thus incurring additional latencies for those requests.}

We hypothesize that by \emph{not} prioritizing PROs and thus giving back more control to the memory request scheduler, we can reduce the execution time and energy consumption overheads of memory-controller-based read disturbance mitigation techniques. \leavevmode{To test this hypothesis, we evaluate state-of-the-art memory-controller-based mitigation techniques
using \leavevmode{\emph{Priority}}
\newpage
\noindent\emph{Scheduling} for PROs, and using a PRO scheduling configuration \leavevmode{(which we call \emph{Insecure}, as described in §~\ref{sec:mot_eval})} in which the memory controller instead places PROs in its read queue. }%

\subsection{Evaluation \leavevmode{Methodology}}
\label{sec:mot_eval}
We evaluate six state-of-the-art read disturbance mitigation techniques: Hydra\leavevmode{\cite{qureshi2022hydra}}, PARA\leavevmode{\cite{kim2014flipping}}, CoMeT\leavevmode{\cite{bostanci2024comet}}, DAPPER\leavevmode{\cite{dapper}}, Graphene\leavevmode{\cite{park2020graphene}}, and ABACuS\leavevmode{\cite{olgun2024abacus}}. For ABACuS, we choose the ABACuS-Big configuration that uses one ABACuS counter per DRAM row ID (i.e., as many counters as there are rows in a bank) to keep track of every sibling row's activation count precisely. \leavevmode{We evaluate the above mitigation techniques using two PRO scheduling configurations:}\footnote{Some mitigation techniques, like \leavevmode{Hydra}~\cite{qureshi2022hydra}, additionally issue read and write requests as part of aggressor tracking, these requests are treated like PROs, meaning that these are prioritized in the {Priority Scheduling} configuration and deprioritized in the {Insecure} configuration.}
\begin{enumerate}
\item \textbf{{Priority Scheduling}:} \leavevmode{The memory controller prioritizes PROs} over all other memory requests \leavevmode{in the memory controller's queues} to ensure timely read disturbance mitigation. \textcolor{black}{This configuration \emph{guarantees} data integrity}. 
\item \textbf{{Insecure}:} \leavevmode{The memory controller places} PROs in \leavevmode{its read queue}, \leavevmode{so} that they are scheduled according to the memory controller's scheduling policy, as opposed to being prioritized over other memory requests. This is an insecure configuration that does \textcolor{black}{\emph{not} guarantee data integrity}.
\end{enumerate}

\leavevmode{We use Ramulator 2.0~}, an open-source cycle-level memory system simulator\leavevmode{\ramulatortwocite{}}. We configure \leavevmode{Ramulator 2.0} using the parameters given in Table~\ref{tab:system_config}.
We simulate a multi-core system with two memory channels for 10 million instructions and continue executing a finished core until all cores have finished to create memory pressure. 
We simulate the evaluated mitigation techniques at a\leavevmode{n \nrh{}} value of 125 using 55 homogeneous memory workloads from
{the SPEC CPU2006\leavevmode{\cite{spec2006}}, SPEC CPU2017\leavevmode{\cite{spec2017}}, TPC\leavevmode{\cite{tpcweb}}, and MediaBench\leavevmode{\cite{fritts2009media}} benchmark suites and \leavevmode{the YCSB benchmark \cite{ycsb}}}. We use the DRFM command to issue PROs. 
 For each mitigation technique, we simulate both PRO \textcolor{black}{scheduling configurations: Priority Scheduling and Insecure}. {We compute the overheads of a \leavevmode{configuration $c$} \leavevmode{normalized to the baseline system (base) without any read disturbance mitigation} as given by Equation~(\ref{eq:overhead-metrics}).}
\begin{equation}
\label{eq:overhead-metrics}
{
\begin{aligned}
\mathrm{Performance\ Overhead}(c)[\%] &=
\left(\frac{\mathrm{cycles}_{c}}{\mathrm{cycles}_{\mathrm{base}}} - 1 \right)\cdot 100, \\[2pt]
\mathrm{DRAM\ Energy\ Overhead}(c)[\%] &=
\left(\frac{\mathrm{energy}_{c}}{\mathrm{energy}_{\mathrm{base}}} - 1 \right)\cdot 100.
\end{aligned}
}
\end{equation}

\leavevmode{We compute the speedup and DRAM energy reduction of the Insecure configuration over the Priority Scheduling configuration as shown in Equation~(\ref{eq:speedup-metrics}).}

\begin{equation}
\label{eq:speedup-metrics}
\small
{
\begin{gathered}
\mathrm{Speedup}[\%]=
\left(
\frac{\mathrm{Cycles}_{\mathrm{Priority Scheduling}}}
     {\mathrm{Cycles}_{\mathrm{Insecure}}}
-1
\right)\cdot 100, \\[2pt]
\mathrm{EnergyRed.}[\%]=
\left(
\frac{\mathrm{Energy}_{\mathrm{Priority Scheduling}}}
     {\mathrm{Energy}_\mathrm{{Insecure}}}
-1
\right)\cdot 100.
\end{gathered}
}
\end{equation}

\subsection{Motivational Results}
\label{sec:mot_res}

\begin{figure}[!ht]
    \centering
    \includegraphics[width=1\linewidth]{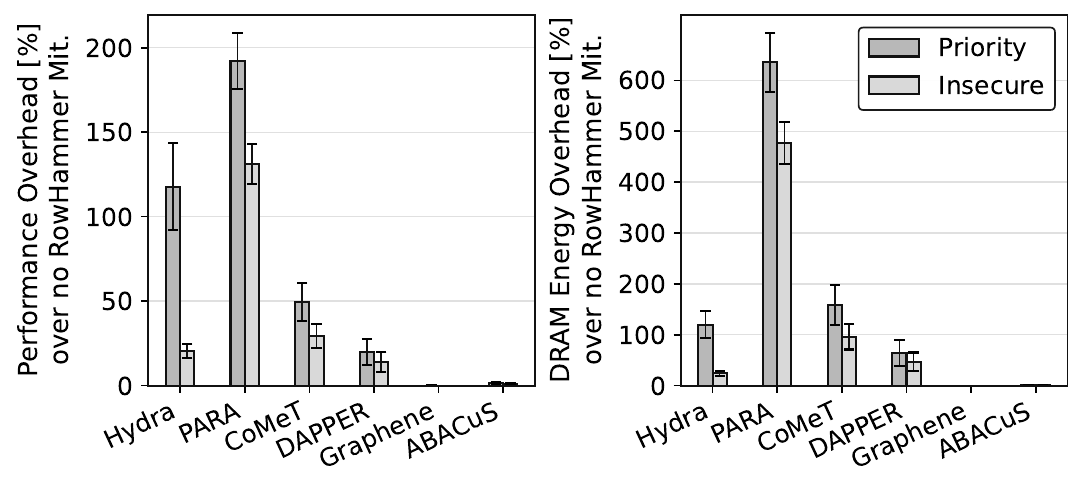}
    \caption{\leavevmode{Performance and DRAM energy overheads of priority scheduling and the insecure configuration over no read disturbance mitigation at \leavevmode{\nrh{}=125}} }
    \label{fig:motivation}
\end{figure}

{Figure~\ref{fig:motivation} shows the average normalized performance and DRAM energy overheads of the Priority Scheduling and the Insecure configurations at an \leavevmode{\nrh{} value} of 125.} 

\leavevmode{Overall, we observe an average speedup of 16.9\% and an average DRAM energy reduction of 18.0\% when using Insecure over Priority Scheduling across all evaluated mitigation techniques.} We observe that Hydra, PARA, CoMeT, and DAPPER experience significantly reduced overheads \leavevmode{\leavevmode{for Insecure} compared to Priority Scheduling. On average across \leavevmode{Hydra, PARA, CoMeT, and DAPPER}, we observe a speedup of 23.0\% and a DRAM energy reduction of 26.9\% for Insecure over Priority Scheduling.}
For ABACuS and Graphene, the benefits of of Insecure are more modest, \leavevmode{with an average speedup and DRAM energy reduction of 0.2\% \textcolor{black}{and 0.2\%, respectively}}. This is because their overheads are quite low to begin with, as \leavevmode{even at a low \nrh{} value} of 125, neither of the mitigation techniques issues many PROs.
{We conclude that deprioritizing PROs leads to substantial overhead reductions for \leavevmode{mitigation techniques that frequently issue PROs (e.g., Hydra~\cite{qureshi2022hydra}, PARA~\cite{kim2014flipping}, CoMeT~\cite{bostanci2024comet}, and Dapper~\cite{dapper})}}.

\leavevmode{We emphasize that the Insecure configuration should \emph{not} 
be implemented to reduce the overheads of read disturbance mitigation techniques. In the Insecure configuration, it is possible for an aggressor row to be activated more times than the PRT before its victim rows are refreshed, which does \emph{not} guarantee data integrity. We ensure data integrity with \X{}~\leavevmode{(§\ref{sec:mech})}, while maintaining most of the performance benefits of the Insecure configuration \leavevmode{(as we show in §~\ref{sec_results})}.}

\section{MORDOR}
\label{sec:mech}

We propose \X{}, a new \leavevmode{Preventive Refresh Operation (PRO)} scheduling mechanism designed to improve performance \leavevmode{and \leavevmode{reduce DRAM} energy consumption} of memory-controller-based read disturbance mitigation techniques, without compromising their data integrity guarantees. \X{} can be implemented alongside \emph{any} memory controller-based technique that uses PROs to mitigate read disturbance. {\X{} allows memory demand requests that do \leavevmode{\emph{not}} activate a row with an outstanding PRO to be scheduled using the underlying memory scheduling policy.}\footnote{In this work, we evaluate \X{} alongside the FR-FCFS memory request scheduler~\textcolor{black}{\cite{rixner2000memory, zuravleff1997controller, mutlu2007stall}} (\secref{sec:results}). However, \X{} is compatible with any memory request scheduler that is able to take the blacklisted status of a PRO into account. We leave it to future work to evaluate \X{} alongside other schedulers \leavevmode{(e.g.,\leavevmode{\cite{moscibroda2007memory,mutlu2009parallelism,kim2010atlas,ghose2013criticality,nesbit2006fair,mutlu2007stall,ebrahimi2011parallel,ausavarungnirun2012staged,usui2016dash,zhao2014firm,subramanian2014blacklisting,subramanian2016bliss,mutlu2008distributed,hur2004adaptive,mukundan2012morse,kim2010thread,subramanian2013mise,subramanian2015application,muralidhara2011reducing,lee2008prefetch,jog2016exploiting,ebrahimi2012fairness,ebrahimi2011prefetch,lee2009improving}})}.}
By tracking DRAM row addresses with outstanding PROs until the PRO is served, \X{} guarantees that \leavevmode{\emph{no}} row is activated past its \leavevmode{Preventive Refresh Threshold (PRT)}, and thus \textcolor{black}{preserves data integrity}.

\subsection{Overview}
\begin{figure}[h]
    \centering
    \includegraphics[width=1\linewidth]{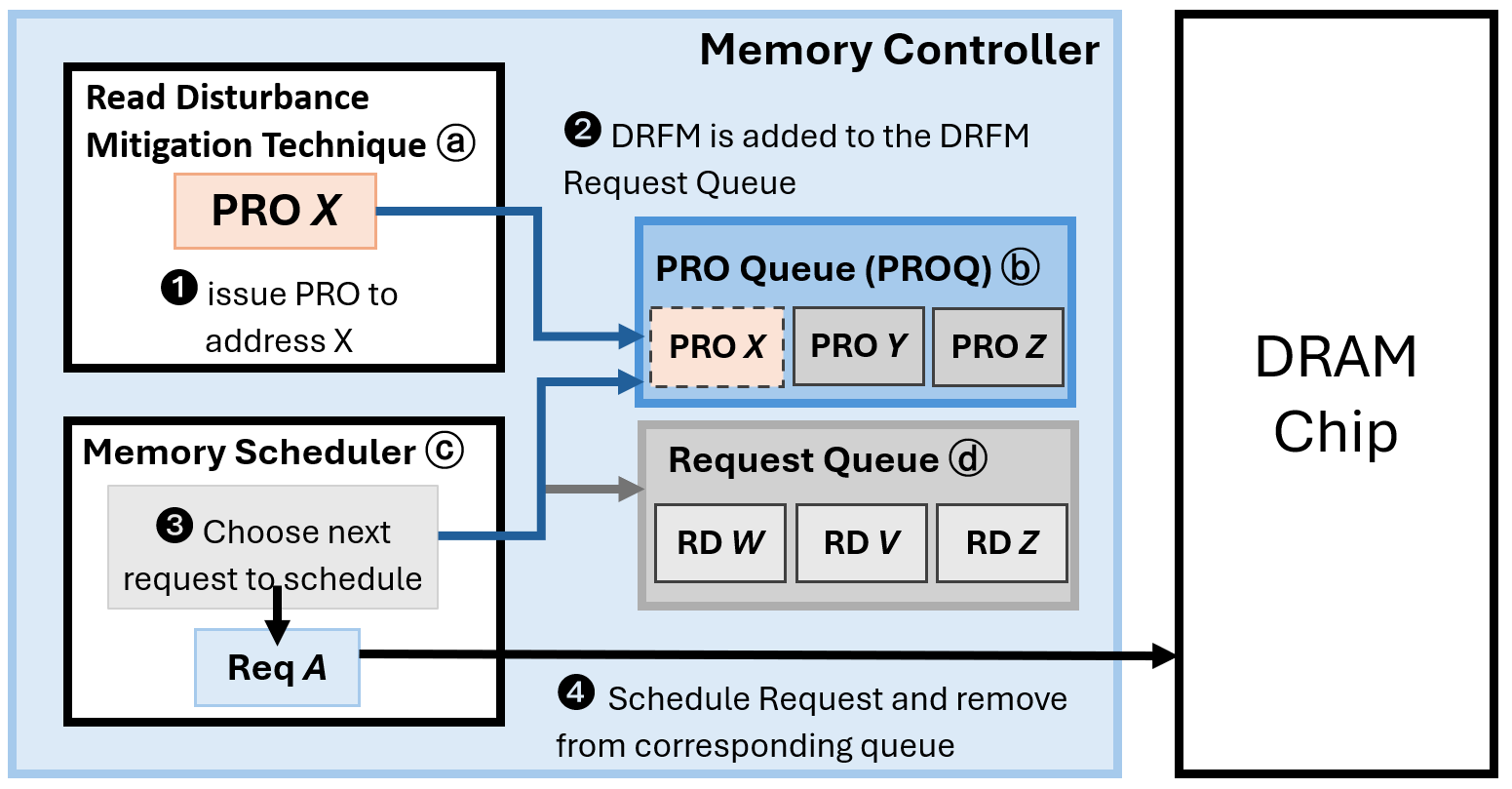}
    \caption{\leavevmode{\leavevmode{\X{} overview}}}
    \label{fig:mech}
\end{figure}

Figure~\ref{fig:mech} shows an overview of \X{}. 
\X{} extends the memory controller with a key component, the Preventive Refresh Operation Queue (PROQ), which stores PROs that have been enqueued by a read disturbance mitigation technique, but not yet issued by the memory controller. \leavevmode{The PROQ acts as a blacklist by \leavevmode{\emph{not}} allowing requests targeting rows with outstanding PROs to proceed.}

\X{} operates in four key steps. First, when a read disturbance mitigation technique~(\circled{a}) identifies a row X as an aggressor row,
\leavevmode{the mitigation technique} issues a PRO to row X (i.e., PRO X) to prevent read-disturbance-induced bitflips in X's victim rows (\ballnumber{1}). Second, \X{} enqueues PRO X in the PROQ~(\circled{b}), (\ballnumber{2}). The PROQ keeps track of all outstanding PROs in the order in which they were enqueued by the mitigation technique. 
\leavevmode{Third, the Memory Request Scheduler~(\circled{c}) checks the Request Queue~(\circled{d}) and the PROQ~(\circled{b}) to find the next request to schedule, and chooses the oldest ready request that does \emph{not} target \leavevmode{\leavevmode{a} row with an outstanding PRO}.}
{When a request \leavevmode{or PRO} is issued by the memory controller, \X{} removes it from the request queue \textcolor{black}{or PROQ, respectively}.}

\subsection{Detailed Design}

\X{}'s PROQ is capable of buffering 32 outstanding PROs. The memory request scheduler is modified to observe both the memory controller's 
request queue and the PROQ when choosing the next request to issue. For a scheduler that follows the First-Ready First-Come First-Served (FR-FCFS) policy, it selects the oldest ready request to a row that does \leavevmode{\emph{not}} have an outstanding PRO.
Since scheduling policies such as FR-FCFS take the age of the request into account when making scheduling decisions, it is important to maintain an age-based ordering among the requests in the PROQ and the request queue.

\leavevmode{W}e add a Request Age Counter (RAC) to each channel of the memory controller which the scheduler uses to compare the relative age of demand requests \leavevmode{in the memory controller's queue} and \leavevmode{the PROQ}. This allows \X{} to preserve the age-ordering semantics of FR-FCFS. Specifically, when multiple ready candidate requests are equivalent under the scheduling policy, the RAC is used to identify the oldest candidate among the request queue and PROQ. The RAC can be implemented as a circular buffer of size 64 (enough to enumerate the requests in the request queue of 32 entries, and the 32 entries of the PROQ). A head and a tail pointer indicate the oldest and youngest requests in the RAC. Each request is extended with a field representing its position in the global request ordering (i.e., the ordering of requests in both the memory controller's request queue and the PROQ).\footnote{Memory controllers that implement separate read and write request queues likely implement a similar way to track the ages of outstanding memory requests for scheduling techniques like FR-FCFS. This approach could also be extended to the PROQ in such a setting. In that case, the counter bits would potentially need to be increased to also track the entries of the PROQ in addition to the request queues.}

The life-cycle of a PRO in a system with \X{} is as follows. 

\noindent\textbf{Enqueueing a PRO. }%
When a row's activation count reaches the PRT, or it is probabilistically selected for mitigation, the read disturbance mitigation technique issues a PRO targeting the row. The memory controller enqueues the PRO in its PROQ, and \leavevmode{the PRO} remains in the PROQ until the memory controller issues it according to the underlying scheduling algorithm. 

\noindent\textbf{Next request scheduling. }
The memory request scheduler chooses the next request to schedule by following the underlying scheduling algorithm and only considering a request as \textit{ready to schedule} if it does \leavevmode{\emph{not}} activate a row that has an outstanding PRO.
\leavevmode{Since the PROQ contains all outstanding PROs, the memory scheduler can cross-reference a request's target row address with the PROQ to find out if that request targets a row with an outstanding PRO.}

\noindent\textbf{Dequeuing a PRO from the PROQ.}
When \leavevmode{the memory controller issues a PRO to an aggressor row, \leavevmode{that row's victims are preventively refreshed}}. \leavevmode{The memory controller removes the PRO} from the PROQ, which allows the memory controller to issue demand memory requests to that row again. 

\leavevmode{\textbf{Use of DRFM.}} To issue PROs in DDR5 in a JEDEC\cite{jedec}-compliant manner, the memory controller can use a command such as DRFM\leavevmode{\cite{jedec}}. \leavevmode{DRFM relies on the memory controller communicating the target address of the DRFM to the DRAM using the DRFM Address Register (DAR). This can be done when the memory controller activates and precharges the target row of the DRFM request~\cite{jedec, taneja2025dream}.} We propose implementing PROs using the DRFM command as follows. \leavevmode{When a mitigation technique issues a PRO to an aggressor row, the memory controller adds that PRO to the PROQ.} Once the memory controller is ready to issue the PRO at the head of the PROQ, it issues an activate and a precharge command followed by a DRFM to that address\leavevmode{, and removes the PRO from the PROQ. The activate and precharge commands store the row's address in the DAR} and ensure that the subsequent DRFM request targets the intended aggressor row. This approach requires reducing the PRT by 1 to account for the additional activate command preceding the DRFM.

\subsection{Security Analysis}
\X{} improves performance and reduces DRAM energy consumption of memory-controller-based read disturbance mitigation techniques by elastically scheduling PROs \emph{without} compromising the data integrity guarantees of the underlying mitigation technique. The security of \X{} is derived from the following three data integrity guarantees\leavevmode{:} 
\begin{enumerate}
    \item {{No} row is activated \leavevmode{\emph{after}} its activation count reaches PRT.}
    \item {The frequency of demand requests to an aggressor row between two PROs remains the same with \X{} as in a setting without \X{}.}
    \item No PROs are ignored due to PROQ capacity. 
\end{enumerate}

\leavevmode{The following sections provide details for each of the security guarantees. }

\textbf{\leavevmode{No Activations Past PRT}.}
\X{} ensures data integrity guarantee (1) as follows.
\leavevmode{\leavevmode{When a row's activation count reaches the PRT, the mitigation technique issues a PRO to that row. \X{} enqueues the PRO into the PROQ, thus blacklisting \leavevmode{the PRO's target row}, and prevents any new activations targeting that row until the PRO is issued and removed from the PROQ. Thus, the row \textit{cannot} be activated during this time, and its activation counter \textit{cannot} exceed the PRT. }}

We assume the underlying read disturbance mitigation technique sets PRT as a safe threshold such that it accounts for the minimum number of activations that cause a bitflip. For a single-sided attack, PRT is calculated based on the number of activations of a single aggressor row. For attacks that use multiple aggressor rows to cause a bitflip in a victim row, PRT is calculated based on the minimum activation count across all aggressor rows. Thus, \X{} ensures that 1)~\leavevmode{\emph{none}} of the \leavevmode{aggressors'} activation counts exceed PRT, and 2)~\leavevmode{\emph{no}} bitflips occur given the underlying read disturbance mitigation technique is secure against bitflips.\footnote{\leavevmode{Note that the PRT can be configured on a per-row basis as described in \textcolor{black}{\cite{yaglikci2024svard, yauglikcci2024enabling}}.}}

{\textbf{Frequency of Demand Requests Between PROs.}}
\leavevmode{\X{} ensures data integrity guarantee (2) as follows.}
{%
Let row $A$ be a row that has reached the PRT and for which a PRO has been enqueued. Enqueuing \leavevmode{the PRO to row \leavevmode{$A$} ensures} that \emph{no} demand request activates row $A$ until its victim rows are refreshed by the PRO.
\leavevmode{\X{} may delay a PRO to serve \emph{only} demand requests that do \emph{not} activate rows with outstanding PROs.} Thus, the number of additional activations to row $A$ \emph{after} the PRO is enqueued by a mitigation technique is zero.}
\leavevmode{We conclude that while a delayed PRO can increase the elapsed time between a PRO being enqueued and the target victim rows being refreshed, it \emph{cannot} increase the number of activations to the aggressor row during that interval. }

\textbf{Preventive Refresh Operation Queue \textcolor{black}{Capacity}.}
\label{sec:PROQ_size}
\leavevmode{\X{} can ensure guarantee 3 with a fixed-size PROQ as follows.}
We observe that with a fixed-sized PROQ, a workload could cause more PROs than the PROQ can store. 
\leavevmode{This would lead to PROs not being placed in the PROQ and the memory controller potentially activating their target rows past the PRT.}
\leavevmode{We consider} the following memory access pattern, \xp{}, engineered to cause a very large number of PROs within a single refresh interval.

\begin{figure}[h]
    \centering
    \includegraphics[width=0.7\linewidth]{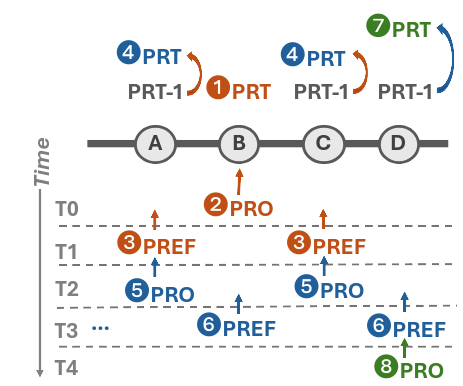}
        \caption{\leavevmode{PROs causing other PROs for rows at PRT-1}}
    \label{fig:pros}
\end{figure}

\leavevmode{\xp{} \leavevmode{consists of} two phases, the \textit{\leavevmode{Prepare} \textcolor{black}{phase}} and the \textit{PRO phase}. During the \leavevmode{Prepare} phase, \xp{} activates as many rows as possible PRT-1 times. \leavevmode{In} the PRO phase, \leavevmode{\xp{} issues activate commands} to increase \leavevmode{those} row's activation count to the PRT and thus cause \leavevmode{a read disturbance mitigation technique to issue a PRO to that row}. 
\leavevmode{In the \leavevmode{Prepare} phase} \xp{} can leverage bank-level parallelism and issue back-to-back activate commands more frequently. During the \leavevmode{PRO} phase, \xp{} leverages the observation \leavevmode{that issuing a PRO to an aggressor row whose victims are at PRT-1 will cause PROs to be issued to \leavevmode{those} victim rows.}
Since the victim rows' \leavevmode{activation counts} are at PRT-1 \leavevmode{when the aggressor receives a PRO}, \leavevmode{preventively refreshing the victim rows}
will raise \leavevmode{their} activation count to PRT and \leavevmode{trigger} a PRO. 
Figure~\ref{fig:pros} illustrates this behavior.
}

\leavevmode{Figure~\ref{fig:pros} shows four neighboring DRAM rows depicted as circles. The y-axis depicts the progression of time in timesteps (top to bottom, T0 to T4) as PROs and preventive refresh (PREF) commands are issued. At timestamp T0, all rows except for row B are at PRT-1 activations, and row B is at PRT activations \ballnumber{1}, \leavevmode{following an activate (ACT) command}. Thus, a mitigation technique issues a PRO to row B, \ballnumber{2}. This PRO preventively refreshes B's neighboring rows, A and C, \ballnumber{3}. This increments the activation count of rows A and C to PRT activations, \ballnumber{4}. As a result, both rows A and C now require a PRO, \ballnumber{5}, which is sent at timestep T2.  The memory controller refreshes C's neighboring rows B and D at timestamp T3, \ballnumber{6}, which increments D's activation count to PRT, \ballnumber{7}. This means that at timestep T4 the memory controller sends a PRO to row D, \ballnumber{8}.}

\leavevmode{We make the key observation that a PRO that is caused by an activate command 
can cause up to \leavevmode{\emph{two}} new PROs, \leavevmode{one PRO for each victim,} (as was the case with the PRO \leavevmode{\ballnumber{2}}), whereas a PRO caused by a PRO can only cause \leavevmode{\emph{one}} new PRO\leavevmode{, \ballnumber{8}} (as we can observe with the PRO sent to row C,  \leavevmode{\ballnumber{5}}), since the victim on the other side (B) has already been \leavevmode{refreshed, \ballnumber{6}}, and its activation count reset.}
\textcolor{black}{ $P_{adv}$ uses this observation to increase the number of simultaneously issued PROs in the PRO phase.}
\leavevmode{To evaluate the PROQ-size requirements of $P_{adv}$, we computationally simulate} the behavior of ACT and PRO commands to estimate the effects of such extremely PRO-intensive memory access patterns. To do so, we simulate a simplified DDR5-based system that focuses exclusively on ACT commands and PROs. We simulate a 32ms refresh window, a request queue of 64 entries, \leavevmode{to match the DRAM configuration evaluated in Section~\ref{sec_results},} with a command frequency of 5~ns for ACTs and 60~ns for PROs \leavevmode{following DDR5 timing parameters ~\cite{jedec2020ddr5, kim2012case}}, at a \leavevmode{PRT} of 125.

We empirically find the duration for the Prepare and PRO phases to maximize the number of PROs during a refresh interval. Based on our results, we estimate the required PROQ size to accommodate all outstanding PROs to be $\approx4700$ entries.
We note that we never observe such high PROQ size requirements for any of the evaluated \leavevmode{workloads.}
To \leavevmode{\leavevmode{guarantee} data integrity in the presence of} adversarial or highly PRO-intensive memory access patterns, we propose the following approach, based on the observation that a \leavevmode{PRO caused by another PRO} can produce at most \emph{one} new PRO. 

\leavevmode{\textbf{PROQ-capacity Promotion Mechanism.} \label{sec:PROQ_cap}
When the PROQ is at risk of reaching its full capacity (i.e., close to 32 PROs are present in a 32-entry PROQ), the memory controller proactively starts prioritizing PROs to reduce the number of PROs in the PROQ to prevent it from overflowing. 
\leavevmode{The PROQ must always have one spare entry available, since issuing a PRO can cause up to \emph{two} new PROs (one for each refreshed victim row).}
Thus, when a 32-entry PROQ reaches 31 entries, the memory controller proactively prioritizes the head PRO in the PROQ.
\leavevmode{Assuming that the head of the PROQ causes \emph{two} new PROs, issuing it} will fill the PROQ to 32 entries.
\leavevmode{Thus, if the PROQ contains 32 PROs, we know that the most recent PROs at the tail of the PROQ were caused by a PRO.}
We leverage the observation that a PRO caused by a PRO can cause \emph{at most one} new PRO. Since we know that the last two PROs in a PROQ with 32 PROs were caused by PROs, we can safely prioritize those, knowing that they will \leavevmode{\emph{not}} create more PROs than the PROQ can store. 
Once the number of PROs in the PROQ is 30 or less, the memory controller returns to issuing PROs based on the underlying request scheduling algorithm, as opposed to prioritizing PROs.}

\leavevmode{\textbf{PRT Adjustment with Mitigation Blast Radius.}} We note that reordering the PROs may lead to a victim \leavevmode{row} refresh caused by a PRO activating \leavevmode{an aggressor row that already has a PRO enqueued in the PROQ. This would lead to that aggressor receiving more than PRT activations, at which point data integrity can no longer be guaranteed.} \leavevmode{To avoid this, the} PRT values should be reduced by two, \leavevmode{when used with a mitigation blast radius of one (i.e., one row refreshed on each side of the aggressor),} \leavevmode{to account for one possible activation from each neighboring row}. \leavevmode{If the read disturbance mitigation blast radius is higher than only one aggressor on each side (i.e., $blast\_radius=1$), the PRT should be decreased by $2\cdot blast\_radius$.} We note that the same overflow pattern is a risk for a fixed-sized priority queue \leavevmode{for the Priority Scheduling configuration}, and the PRT values should be set accordingly.

\subsection{Integration with PRAC}
\label{sec:prac}
\leavevmode{In addition to memory controller-based techniques, \X{} can} be used for performance improvement in a setting that implements \leavevmode{in DRAM} Per-Row Activation Counting (PRAC)~\textcolor{black}{\cite{canpolat2025chronus, canpolat2024understanding, jedec2020ddr5}}. 
\leavevmode{PRAC implements an activation counter for each DRAM row, and thus accurately measures the activation counts of \emph{all} rows. When a row’s activation count reaches the PRT, the DRAM chip asserts an alert back-off signal which forces the memory controller to issue a PRO.}
\leavevmode{Similarly to the Priority Scheduling configuration, raising the alert back-off signal}
preempts the schedule generated by the memory controller. 

\leavevmode{This can lead to increased latencies for demand memory requests.}
{\leavevmode{\X{} can be used alongside} PRAC-based mitigation techniques to alleviate demand memory request latencies \leavevmode{given the aggressor row information (e.g., by tracking recently activated rows in the memory controller)}. Instead of pausing 
\leavevmode{demand memory} requests when the alert back-off signal is received, the memory controller \leavevmode{\leavevmode{enqueues a PRO}
\leavevmode{into its PROQ}.}
\leavevmode{This way} the memory controller can continue to send requests to \leavevmode{rows that do \emph{not} have outstanding PROs} in a timely manner \textcolor{black}{and demand memory requests experience less delay}. Instead of the DRAM performing the victim row mitigation directly, \leavevmode{the memory controller schedules the PRO} according to the memory controller's scheduling policy.}

\section{Hardware Implementation}
\X{} is implemented in the memory controller and does \leavevmode{\emph{not}} require modifications to any other part of the system.
To enable \X{}, two key components must be added to the memory controller: the Preventive Refresh Operation Queue (PROQ) and the Request Age Counter (RAC). The PROQ keeps track of outstanding Preventive Refresh Operations (PROs), while the RAC tracks the request ordering for the PROs in the PROQ and the requests in the memory controller's request queue. 

\leavevmode{For a memory controller with a 32-entry request queue \X{} can be implemented as follows.}
Each entry of the PROQ stores the 24-bit address of the PRO along with 6 bits of ordering metadata for the request ordering given by the RAC. The total storage required by the PROQ is 120 bytes. For a 32-entry PROQ, we require a 64-entry RAC (which gives the ordering for 32 PROs in the PROQ and a 32-entry request queue). Two byte-sized pointers indicate the head and tail (oldest and youngest) entries of the RAC. Overall, the RAC requires a total of 50 bytes. The total storage required by \X{} constitutes only 170 bytes.
We implement the PROQ using a content-addressable memory (CAM) array, since PROQ is often queried by the memory request scheduler to identify the next request to schedule.

\section{Evaluation Methodology}
\label{sec:meth}
We evaluate the performance and DRAM energy overheads of \X{} when used with the following six state-of-the-art read disturbance mitigation techniques: Hydra~\cite{qureshi2022hydra}, PARA~\cite{kim2014flipping}, CoMeT~\cite{bostanci2024comet}%
, DAPPER~\cite{dapper}%
, Graphene~\cite{park2020graphene}, and ABACuS~\cite{olgun2024abacus} \leavevmode{(described in detail in §\ref{sec:mech_background})} at four \leavevmode{\nrh{} values} of 125, 250, 500, and 1000. For our evaluation, we use Ramulator 2.0, an open-source cycle-level memory system simulator~\textcolor{black}{\ramulatortwocite{}}. 
\leavevmode{Table~\ref{tab:system_config} shows the simulated system configuration.}

\vspace{10pt}
\begin{table}[h]
\centering
\caption{Simulated system configuration }
    \resizebox{0.9\linewidth}{!}{
\begin{tabularx}{\linewidth}{>{\bfseries}p{0.25\linewidth}|X}
\hline
\textbf{Processor} & 8 cores, 4.3GHz clock frequency, 4-wide instruction issue, 128-entry reorder buffer\\
\hline
\textbf{DRAM} & DDR5\leavevmode{\cite{jedec}}, 2 channels, 2 ranks/channel, 8 bank groups, 2 banks/bank group, 64K rows/bank, mitigation blast radius 1, tRP=15.0ns, tRCD=15.0ns, tRC=47.5ns, tRAS=32.5ns\\
\hline
\textbf{Memory Controller} & Address Mapping: MOP4CLXOR\cite{kaseridis2011minimalistic}, Row Policy: Open Row Policy, 32-entry read and write queues\\
\hline
\textbf{Last-Level Cache} & 64-byte cache lines, 8 set associative, 2MB capacity/core, latency: 47 cycles\\
\hline
\end{tabularx}}
\label{tab:system_config}
\end{table}
\vspace{10pt}

We evaluate 55 homogeneous workloads from {four} benchmark suites: SPEC CPU2006~\cite{spec2006}, SPEC CPU 2017~\cite{spec2017}, TPC~\cite{tpcweb}, {and }MediaBench~\cite{fritts2009media}, {as well as workloads from} YCSB~\cite{ycsb}.
{We categorize the tested workloads based on the number of PROs that they issue.}
We define the PRO-intensity of a workload as the total number of PROs that the evaluated mitigation techniques issue when executing that workload for 10 million instructions in the 8-core setting.
Table~\ref{table:workloads} shows the classification of the evaluated memory traces by \leavevmode{the average number of PROs issued per kilo-instruction (PROPKI)}.\footnote{For Hydra, we include mitigation-related read and write requests in this count. For ABACuS, one DRFMab\leavevmode{\cite{jedec}} request is counted as one PRO.}

We use the DRFM command with a blast radius of one (meaning that one victim row is refreshed on either side of the aggressor row for each DRFM) to issue PROs. 
We use DRFMsb\leavevmode{~\cite{jedec}} for all tested mitigation techniques except for ABACuS~\textcolor{black}{\cite{olgun2024abacus}}, where we use DRFMab\leavevmode{~\cite{jedec}} due to the coarser granularity at which it issues mitigations.\footnote{Our main evaluation uses the DRFM latency provided by the evaluated DDR5
timing model~\textcolor{black}{\cite{jedec, luo2023ramulator, ramulator2github}} and does \leavevmode{\emph{not}} explicitly model the ACT used to populate the
DRFM address register. Section~\ref{sec:drfm-act-sensitivity} evaluates a
conservative model that charges the corresponding address-setup latency
only to MORDOR. }

Table \ref{tab:config_parameters} shows the configuration parameters used for Hydra, PARA, CoMeT, DAPPER, Graphene, and ABACuS in our evaluation for varying PRT values. To simulate ABACuS, we implement the ABACuS-Big configuration\leavevmode{~\cite{olgun2024abacus}}. To ensure that PROQ-capacity-caused PRO \leavevmode{promotions (§\ref{sec:PROQ_size})} \textcolor{black}{do \leavevmode{\emph{not}} affect data integrity}, we reduce the \nrh{} value by 2 (\leavevmode{i.e.,} $2\cdot blast\_radius$) when using \X{}. \leavevmode{This way \X{} can guarantee data integrity at a fixed PROQ size of 32 entries.}

\begin{table}[h]
\small
\centering
\caption{\leavevmode{Evaluated single-core workloads \leavevmode{(PROPKI: PROs per kilo-instruction)}}}
\begin{tabular}{|l||l|}
\hline
\leavevmode{\textbf{PROPKI}} & \leavevmode{\textbf{Workloads}} \\ \hline \hline
\begin{tabular}[c]{@{}l@{}}\leavevmode{[3.13+)} \\ \leavevmode{(High)}\end{tabular} & \begin{tabular}[c]{@{}l@{}}
GUPS, 549.fotonik3d, stream, tpcc64 557.xz,\\
470.lbm, 429.mcf, 437.leslie3d, 462.libquantum, \\
450.soplex, 482.sphinx3, 483.xalancbmk, tpch17,\\ 
520.omnetpp, 436.cactusADM, 505.mcf, 510.parest, \\
459.GemsFDTD, tpch2, 
471.omnetpp, 473.astar\\
\end{tabular} \\ \hline
\begin{tabular}[c]{@{}l@{}}\leavevmode{[0.06, 3.13)}\\ \leavevmode{(Medium)}\end{tabular} & \begin{tabular}[c]{@{}l@{}}
 507.cactuBSSN,ycsb\_aserver, wc\_8443, wc\_map0,\\
 ycsb\_bserver, ycsb\_cserver, ycsb\_dserver, \\
 456.hmmer, ycsb\_eserver, h264\_encode,\\
 ycsb\_abgsave, grep\_map0, tpch6, 401.bzip2,\\ 
 500.perlbench \\
\end{tabular} \\ \hline
\begin{tabular}[c]{@{}l@{}}\leavevmode{[0, 0.06)} \\ \leavevmode{(Low)}\end{tabular} & \begin{tabular}[c]{@{}l@{}}
464.h264ref, 523.xalancbmk, 526.blender, 502.gcc,\\
508.namd, 435.gromacs, 445.gobmk, 544.nab, \\
458.sjeng, 531.deepsjeng, jp2\_encode, 481.wrf, \\
447.dealII, 525.x264, 403.gcc, 444.namd, \\
541.leela, 511.povray, 538.imagick
\end{tabular} \\ \hline
\end{tabular}
\label{table:workloads}
\end{table}

\vspace{10pt}
\begin{table}[h]
\centering
\caption{Configuration parameters and their values for various mitigation techniques}
\large
\setlength{\tabcolsep}{2pt}
\newcommand{\twoline}[2]{\begin{tabular}[t]{@{}l@{}}#1\\#2\end{tabular}}
\resizebox{\linewidth}{!}{
\begin{tabular}{|l|l|*{4}{>{\centering\arraybackslash}p{1.3cm}|}}
\hline
\textbf{Mechanism} &
\textbf{\twoline{Configuration}{Parameter}} &
\multicolumn{4}{c|}{\textbf{Value (Priority / MORDOR)}} \\ \hline 
\hline

All Mechanisms &
\leavevmode{\twoline{RowHammer Threshold}{\leavevmode{(\nrh{})}}}&
125 / 123 & 250 / 248 & 500 / 498 & 1000 / 998 \\ \hline
\hline

\multirow{3}{*}{Graphene~\textcolor{black}{\cite{park2020graphene}}}
& Number of table entries
& 10,377/ 10,546
& 5,189/ 5,231
& 2,595 / 2,605
& 1,298 / 1,300 \\ \cline{2-6}
& \twoline{Threshold of}{aggressor tracking}
& 63 / 62
& 125 / 124
& 250 / 249
& 500 / 499 \\ \cline{2-6}
& Reset window
& \multicolumn{4}{c|}{32ms} \\ \hline
\hline

\multirow{1}{*}{PARA~\textcolor{black}{\cite{kim2014flipping}}}
& Probability threshold
& 0.241 / 0.245
& 0.129 / 0.130
& 0.067 / 0.067
& 0.034 / 0.034 \\ \hline
\hline

\multirow{6}{*}{Hydra~\textcolor{black}{\cite{qureshi2022hydra}}}
& Row group size
& \multicolumn{4}{c|}{128 rows} \\ \cline{2-6}
& Row count table entry size
& \multicolumn{3}{c|}{1B} & 2B \\ \cline{2-6}
& Row count cache size
& \multicolumn{4}{c|}{4K entries per DRAM rank} \\ \cline{2-6}
& \twoline{Group count table}{threshold}
& 50 / 49
& 100 / 99
& 200 / 199
& 400 / 399 \\ \cline{2-6}
& Tracking threshold
& 64 / 63
& 125 / 124
& 250 / 249
& 500 / 499 \\ \cline{2-6}
& Periodic reset
& \multicolumn{4}{c|}{32ms} \\ \hline
\hline

\multirow{2}{*}{ABACuS~\textcolor{black}{\cite{olgun2024abacus}}}
& \twoline{Preventive refresh}{threshold}
& 62 / 61
& 125 / 124
& 250 / 249
& 500 / 499 \\ \cline{2-6}
& Refresh cycle threshold
& 60 / 59
& 123 / 122
& 248 / 247
& 498 / 497 \\ \hline
\hline

\multirow{4}{*}{CoMeT~\textcolor{black}{\cite{bostanci2024comet}}}
& Activation threshold
& 32 / 31
& 63 / 62
& 125 / 125
& 250 / 250 \\ \cline{2-6}
& Number of hashes
& \multicolumn{4}{c|}{4} \\ \cline{2-6}
& \twoline{Number of counters}{per hash}
& \multicolumn{4}{c|}{512} \\ \cline{2-6}
& RAT size
& \multicolumn{4}{c|}{128} \\ \hline
\hline

\multirow{2}{*}{DAPPER~\textcolor{black}{\cite{dapper}}}
& Activation threshold
& 62 / 61
& 125 / 124
& 250 / 249
& 500 / 499 \\ \cline{2-6}
& Reset window
& \multicolumn{4}{c|}{32ms} \\
\hline
\end{tabular}%
}
\label{tab:config_parameters}
\end{table}

\section{Results}
\leavevmode{We evaluate the execution time and {DRAM} energy consumption of {read disturbance mitigation techniques when used {with}} \X{} compared to {when used {with}} {the Priority Scheduling configuration, where the memory controller prioritizes} {Preventive Refresh Operations (PROs)} {over all other requests in the memory controller's queues}. To this end, we simulate {six} state-of-the-art read disturbance mitigation techniques {(described in §\ref{sec:mech_background})} at various {\nrh{} values}.}
\label{sec_results}
\subsection{Speedup and DRAM Energy Reduction}

\begin{figure}[h]
    \centering
    \includegraphics[width=01.0\linewidth]{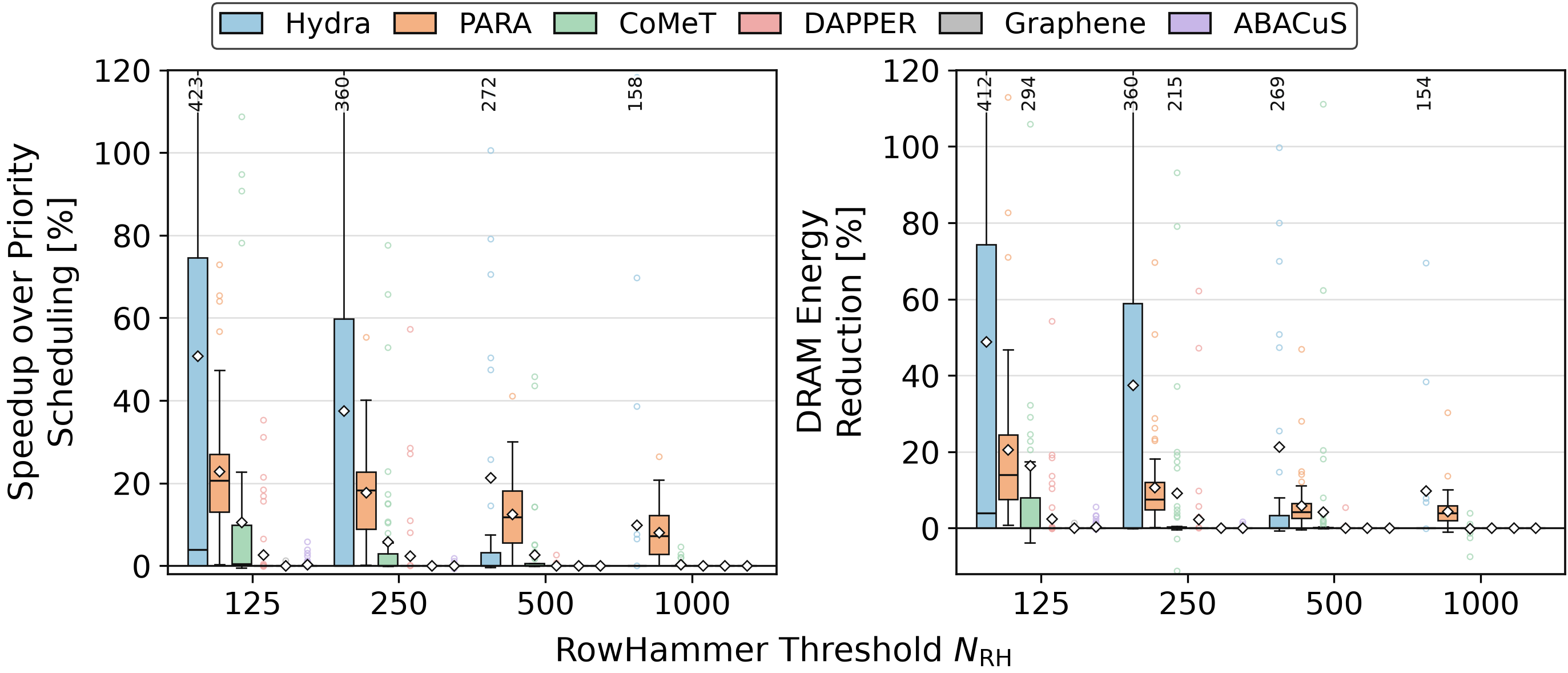}
    \caption{\leavevmode{Speedup and DRAM energy reduction from \X{} compared to \leavevmode{P}riority \leavevmode{S}cheduling}}
    \label{fig:main_res}
\end{figure}

{Figure~\ref{fig:main_res} reports the \leavevmode{speedup and DRAM energy reduction of \X{} over} Priority Scheduling. \leavevmode{We compute the speedup and energy reduction according to Equation~\ref{eq:speedup-metrics}.}}

\begin{figure*}[t]
    \centering
    \includegraphics[width=1.0\linewidth]{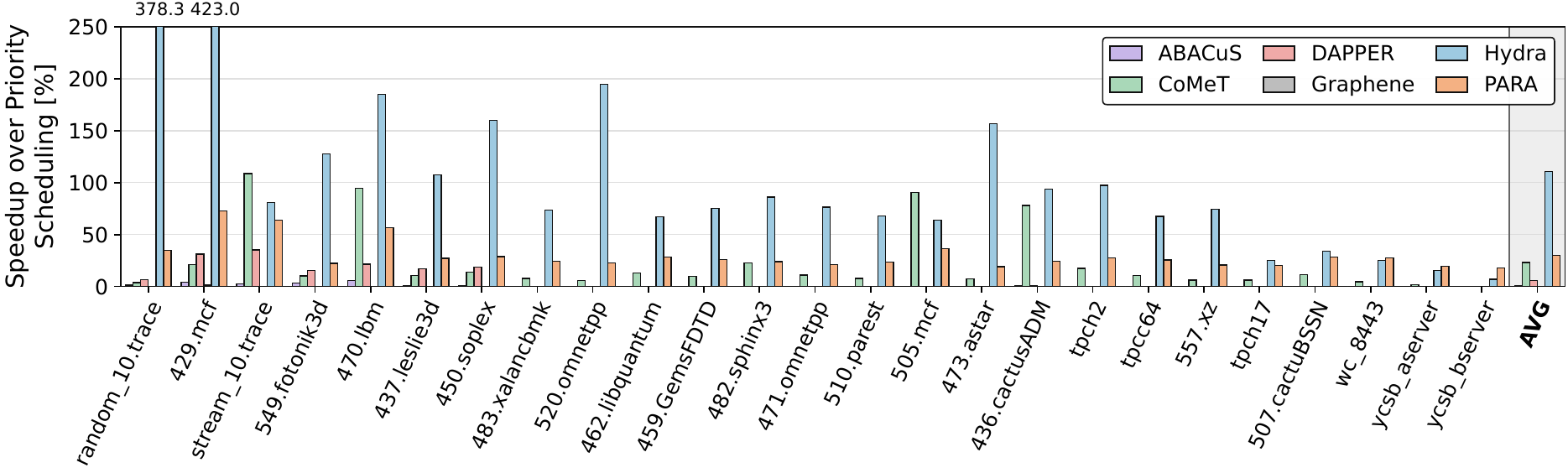}
    \caption{\leavevmode{Per-trace speedup and DRAM energy reduction from \X{} compared to priority scheduling PROs for the 25 most PRO-intensive workloads at \nrh{}=125}}
    \label{fig:per_trace}
\end{figure*}

\label{sec:results}
First, we observe an average (maximum) speedup of \leavevmode{50.8}\% (\leavevmode{423.0}\%), \leavevmode{22.8}\% (\leavevmode{72.9}\%), \leavevmode{10.5}\% (\leavevmode{108.7}\%), \leavevmode{2.7}\% (\leavevmode{35.3}\%), \leavevmode{0.0}\% (\leavevmode{1.2}\%), and \leavevmode{0.3}\% (\leavevmode{5.8}\%) for Hydra, PARA, CoMeT, DAPPER, Graphene, and ABACuS, respectively, when using \X{} at \leavevmode{\nrh{} 125}. \leavevmode{W}e observe an average (maximum) energy consumption reduction of \leavevmode{48.9}\% (\leavevmode{411.5}\%), \leavevmode{20.5}\% (\leavevmode{112.9}\%), \leavevmode{16.3}\% (\leavevmode{293.8}\%), \leavevmode{2.4}\% (\leavevmode{54.2}\%), \leavevmode{0.0}\% (\leavevmode{1.5}\%), and \leavevmode{0.4}\% (\leavevmode{5.6}\%) for Hydra, PARA, CoMeT, DAPPER, Graphene, and ABACuS, respectively \leavevmode{at \leavevmode{\nrh{}$=$125}}.

Second, across the evaluated mitigation techniques, we observe more pronounced benefits for more PRO-intensive mitigation techniques \leavevmode{(i.e., mitigation techniques that frequently issue PROs)} like Hydra, PARA, and CoMeT. PRO-intensive techniques create more interference from PROs with other memory requests when PROs are prioritized. \leavevmode{Hydra is a strong example of this, since it issues both PROs and tracking-related read and write requests that the priority
scheduling approach schedules aggressively.
MORDOR, on the other hand, gives demand memory requests the opportunity to proceed \textcolor{black}{without further delay}.}

Fourth, for ABACuS and Graphene\leavevmode{,} we do \leavevmode{\emph{not}} observe significant improvements from \X{} on average across \leavevmode{workloads.} We attribute this to the fact that these two mitigation techniques are the least PRO-intensive among the evaluated techniques. DAPPER, Graphene, and ABACuS have significantly fewer workloads that trigger PROs,
with Graphene being by far the least PRO-intensive.

Figure~\ref{fig:per_trace} shows the speedup provided by \X{} over
\leavevmode{Priority Scheduling} for Hydra, PARA, CoMeT, DAPPER, Graphene, and ABACuS.
The x-axis shows the 25 most PRO-intensive workloads ordered by PRO intensity.
First, we observe that mitigation techniques behave differently alongside \X{} for different workloads. Hydra experiences large but variable improvements, whereas PARA observes smaller but more consistent benefits. \leavevmode{We attribute this to the fact that Hydra's tracking behavior is more access-pattern-specific, whereas PARA issues PROs at a more steady rate, which depends on PARA's probability threshold. CoMeT also experiences variable improvements, likely due to its workload-specific and bursty mitigation strategy.} Second, we observe that PRO traffic does \leavevmode{\emph{not}} solely determine the benefit of implementing \X{}, as performance improvement is \leavevmode{\emph{not}} monotonic from left to right, even within one mitigation technique. This suggests that the way traffic interacts with demand requests, such as timing, locality, and whether a PRO is on the critical path of a demand request, also plays a role in \textcolor{black}{\X{}'s performance}. 
\leavevmode{We conclude that \X{} improves performance and reduces DRAM energy consumption, especially for low \nrh{} values and for mitigation techniques that frequently issue PROs.}

\subsection{Hardware Overhead Analysis}
{\X{} is implemented entirely in the memory controller and does \leavevmode{\emph{not}} require any modification to the DRAM chip. We implement \X{} in Verilog HDL and use OpenROAD\cite{OpenROAD2025} to estimate \X{}’s PROQ area, power consumption, and PROQ access latency overheads.}
{To put the area overhead into perspective, we compare the estimated area of MORDOR against a 698mm$^2$ Cascade Lake-SP XCC die\cite{thomaskrenn_cascade_lake}. We evaluate \X{} with 6 PROQs, one for each channel on the CPU die. We synthesize our design at 45nm and scale the resulting hardware estimates to 14nm using DeepScaleTool\cite{deepergithub}.\footnote{At the time of writing, \leavevmode{there are no open source PDKs at 14nm, which is why we synthesize using nangate45 and scale the result.}} Table~\ref{tab:mordor-hw-overhead} reports the storage, area, static power, dynamic power, and PROQ access latency of \X{} for various PROQ sizes.}

\begin{table}[h]
\centering
\caption{{Hardware overhead of \X{}.
Area is compared against a 698mm$^2$ Cascade Lake-SP XCC die with 6 memory channels at 14nm.}}
\label{tab:mordor-hw-overhead}
\scriptsize
\setlength{\tabcolsep}{7pt}
\begin{tabular}{r||rrrrrr}
\hline
\textbf{PROQ} &
\textbf{PROQ} &
\textbf{Area} &
\textbf{\% CPU} &
\textbf{Static} &
\textbf{Dyn.} &
\textbf{Access} 
\\
\textbf{entries} &
\textbf{CAM KB} &
\textbf{mm$^2$} &
&
\textbf{mW} &
\textbf{mW} &
\textbf{lat. ns}\\
\hline
\textbf{32} & \textbf{0.72} & \textbf{0.058} & \textbf{0.008}  & \textbf{13.60} & \textbf{145.14} & \textbf{0.51} \\
48 & 1.08 & 0.089 & 0.012 & 22.63 & 191.65  & 0.50 \\
64 & 1.44 & 0.118 & 0.017 & 28.71 & 255.88 & 0.54 \\
78 & 1.74 & 0.144 & 0.020 & 46.18 & 311.26  & 0.67 \\
\hline
\end{tabular}
\end{table}

\begin{figure}[h]
    \centering
    \includegraphics[width=0.9\linewidth]{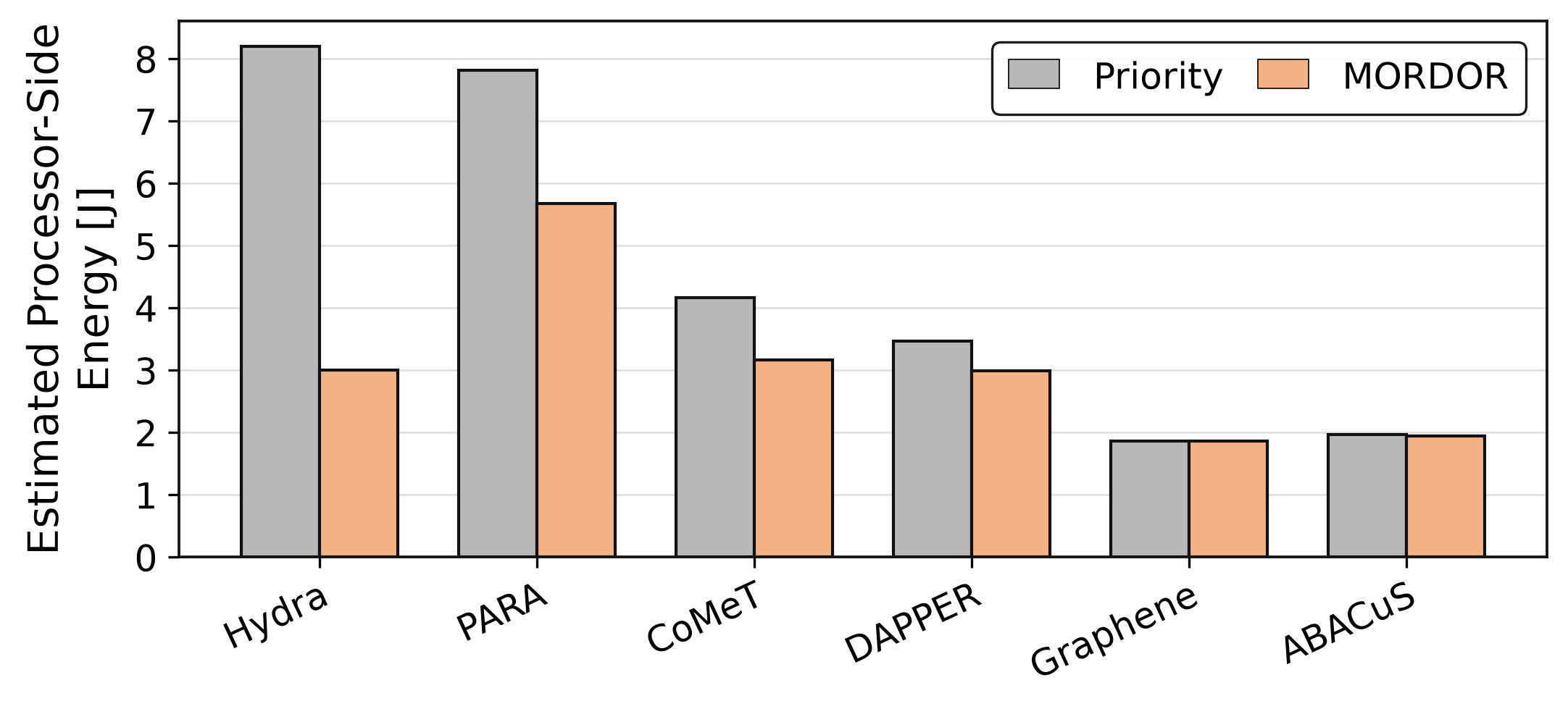}
    \caption{\leavevmode{Per-mechanism processor-side average energy for \X{} and Priority Scheduling at \leavevmode{\nrh{}}=125}}
    \label{fig:power_overheads}
\end{figure}

{\textbf{\leavevmode{PROQ} Energy Overheads.}}
{We estimate the CPU-level power and energy impact of \X{}. We use the Intel Xeon 6980P Processor\cite{intel_xeon_6980p_specs} with a \leavevmode{TDP} of 500W and a base frequency of 2GHz as a coarse processor-power reference and add \X{}'s hardware power to this baseline.}
Figure~\ref{fig:power_overheads} shows the estimated energy of \X{} (implemented with a PROQ of 32 entries) compared to Priority Scheduling across the evaluated read disturbance mitigation techniques. The energy reported for Priority Scheduling includes only CPU energy, while MORDOR energy includes both CPU energy and \X{}'s CAM hardware energy reported by OpenRoad. To obtain processor-side energy for \leavevmode{\X{} and Priority Scheduling}, we use the average simulated execution time across all evaluated workloads. For each of the evaluated mitigation techniques, we use the average cycle count for \X{} and Priority Scheduling at \leavevmode{\nrh{}=125}.

\leavevmode{\leavevmode{With \X{},} the average performance of the evaluated workloads improves, reducing the average execution time. As a result, even with the additional energy required for PROQ management, \X{} reduces overall energy consumption compared to a scheduling approach that prioritizes PROs.}

{\textbf{PROQ-Lookup Latency Overheads.}}
A prior work\cite{subramanian2016bliss} reports the critical path latency of the FR-FCFS scheduler as 1ns for 32nm, which is 0.86ns at 14nm, scaled using DeepScaleTool. We observe that for PROQ sizes of 32, 48, 64, and 78, the PROQ lookup fits into the critical path and does \leavevmode{\emph{not}} incur additional latency overhead, as it can be overlapped with \leavevmode{FR-FCFS'} other readiness checks.

\subsection{Request Latency Analysis}
\begin{figure}[h]
    \centering
    \includegraphics[width=1\linewidth]{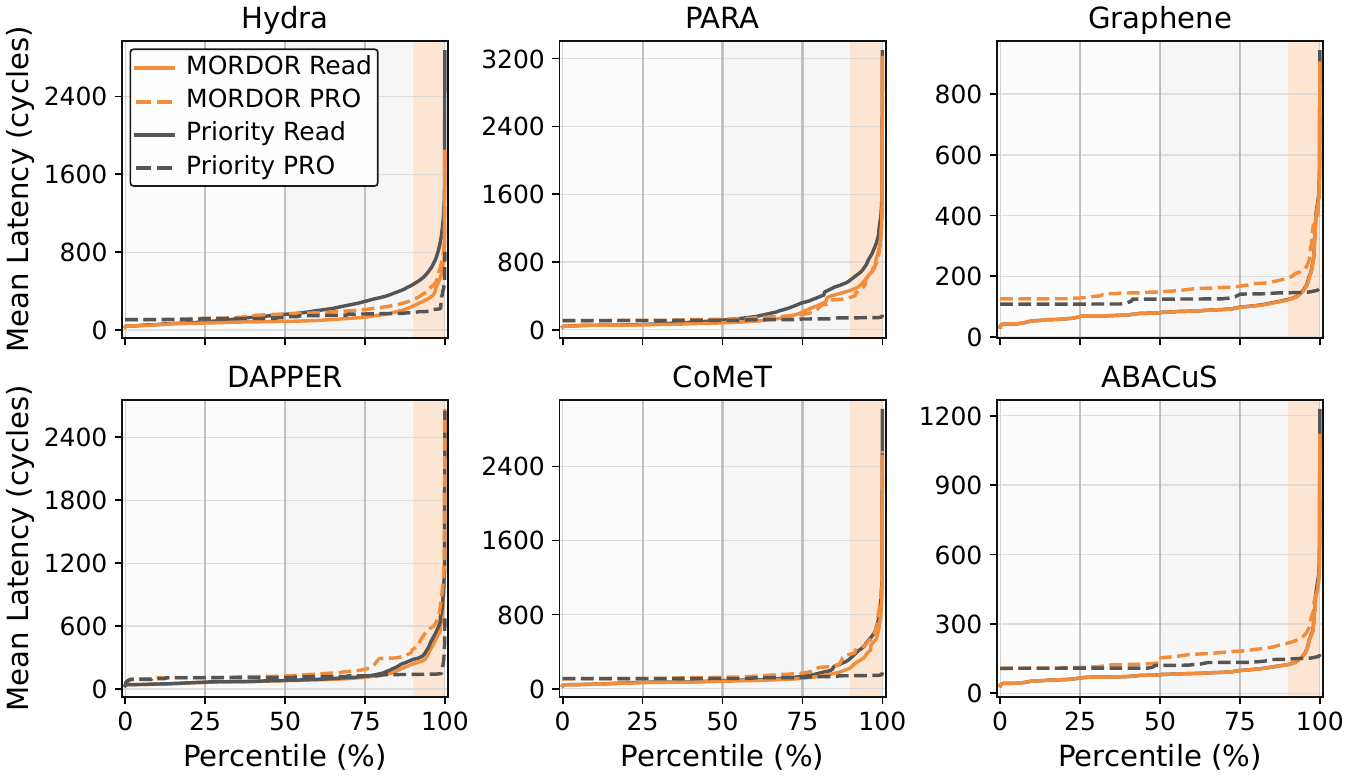}
    \caption{Average read and PRO latencies when using \X{} and Priority Scheduling PROs for 5 PRO-intensive workloads}
    \label{fig:lat_pro_int}
\end{figure}

\leavevmode{We evaluate the impact of \X{} on the memory request latency to better understand the source \leavevmode{\X{}'s speedup over Priority Scheduling}. 
\leavevmode{To measure request latencies, we use Ramulator 2.0's cycle values from when the request is issued until the cycle where data is returned for read requests, or the DRFM command is issued for PROs.}
Figure~\ref{fig:lat_pro_int} shows the average memory request latency percentiles of read requests and PROs for \X{} across the tested read disturbance mitigation techniques, at \leavevmode{\nrh{}=125}, for five PRO-intensive memory workloads, in the single-core setting. The x-axis shows the percentile values, while the y-axis shows the corresponding read request latency in cycles. We first compute each workload’s latency-percentile curve and then report the arithmetic mean of the corresponding percentile values across the five workloads.
For example, the 90th percentile point \leavevmode{(highlighted in orange)} on the x-axis shows that 90\% of the average request latencies are below the latency value on the y-axis.}

We observe that \X{} significantly reduces the read latencies, while increasing the PRO latencies for each tested mitigation technique. \leavevmode{With \X{}, fewer reads are blocked waiting for PROs to complete, because reads to rows \emph{without} outstanding PROs can proceed. This leads to an average latency reduction for read requests.} While the PRO latency increases with \X{}, the total execution time of a memory trace is reduced since a mitigation technique produces significantly fewer PROs than there are reads for a given workload. 
The flatter shape of the latency curve for read requests after the 25th-50th percentiles indicates that fewer read requests experience higher latencies when using \X{}. This is most visible in Hydra, where we can see a pronounced gap between the read-request latencies for \X{} and priority scheduling above the 50th percentile. A similar gap, but to a lesser magnitude, can be observed for PARA, CoMeT, and DAPPER. 

We observe an average reduction in tail latency for the 99th percentile of read requests for all mitigation techniques, \leavevmode{as can be seen with the solid orange line (reads when using \X{}) being shorter than the solid gray line (reads when using Priority Scheduling)}. 
For some individual PRO-intensive workloads, we observe a slight tail-latency increase for read requests in the 99-100th percentile when using \X{}. This is consistent with the case where a read request is waiting on a PRO towards the end of a relatively full PROQ, meaning that it has to potentially wait for multiple PROs to be \leavevmode{issued from the PROQ}. \leavevmode{We observe increased tail latencies only for a subset of the evaluated workloads.} \leavevmode{We therefore attribute the increase in tail latencies to workload-dependent contention between demand requests and queued PROs, rather than to a systematic latency penalty introduced by \X{}.}

\begin{figure}[h]
    \centering
    \includegraphics[width=1.0\linewidth]{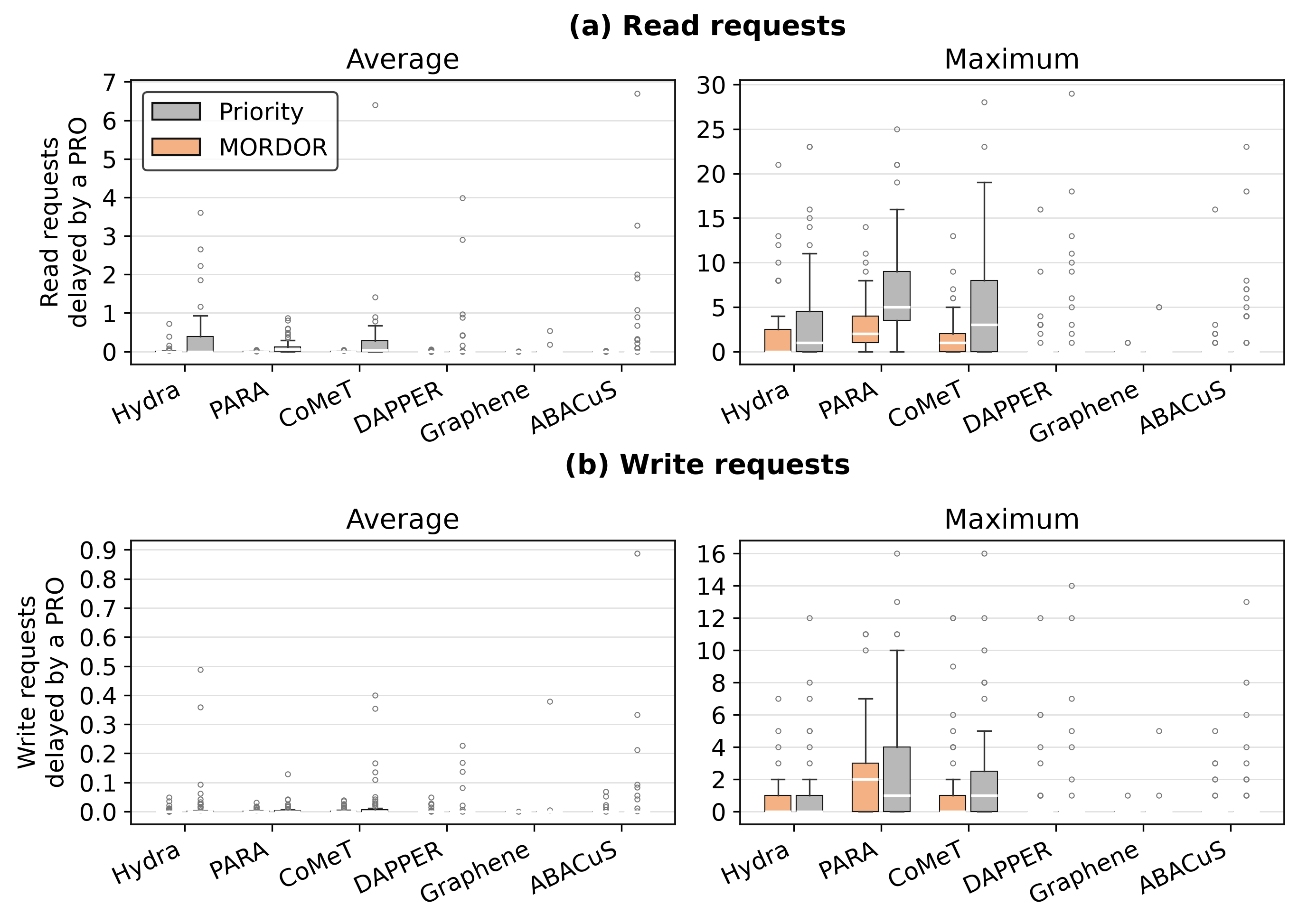}
    \caption{\leavevmode{Average and maximum number of ready read and write requests delayed by a PRO at \leavevmode{\nrh{}}=125}}
    \label{fig:blocked_by_pro}
\end{figure}

\leavevmode{T}he observed demand-request latency reduction comes from the fact that \X{} takes PROs off the critical path of demand requests.
\leavevmode{Figure~\ref{fig:blocked_by_pro} shows the average and maximum number of read (a) and write (b)} requests that are delayed by a PRO \leavevmode{when using \X{} and when using Priority Scheduling}. A request is counted as delayed by a PRO if it is ready to be scheduled, but a PRO is chosen to be scheduled instead by the memory controller's scheduling algorithm.

We observe that \X{} significantly reduces both the average and the maximum number of demand memory requests delayed by PROs across the evaluated mitigation techniques. This reduction largely comes from the fact that with \X{}, significantly fewer PROs end up on the critical path of another request since the memory controller schedules PROs following FR-FCFS and consequently prefers older, ready demand requests when picking the next request to schedule. 
\leavevmode{Specifically, we observe that the average number of read requests delayed by PROs reduces from 0.21 to $\ll 0.01$ when using \X{} compared to Priority Scheduling. The maximum number of read requests delayed by PROs reduces from 3.30 to 1.36. Similarly, the average number of write requests delayed by PROs reduces from 0.02 to $\ll 0.01$, and the maximum number of write requests delayed by PROs reduces from 1.25 to 0.88 when using \X{} compared to Priority Scheduling.}

\leavevmode{We conclude that taking PROs off the critical path of demand requests \leavevmode{\emph{not}} targeting rows with outstanding PROs reduces the number of demand requests that a PRO can block. This allows demand requests to be scheduled sooner, reducing their average latency, and improves average \leavevmode{workload performance}.}

\subsection{{Area vs\leavevmode{.} Performance Tradeoff Analysis}}

\begin{figure}[H]
    \centering
    \includegraphics[width=1.0\linewidth]{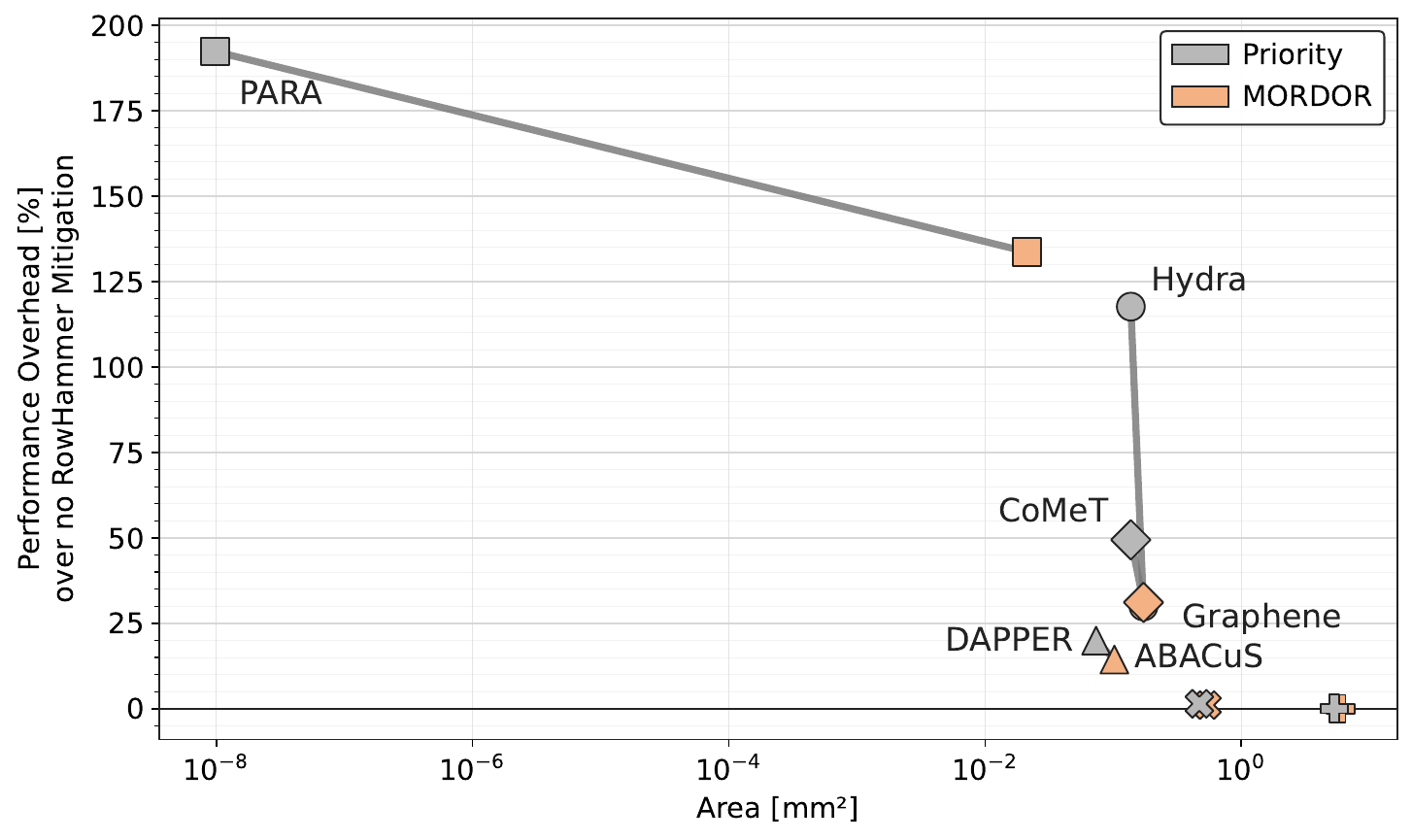}
    \caption{\leavevmode{Area vs\leavevmode{.} performance overhead over no mitigation when using \X{} and Priority Scheduling evaluated at \leavevmode{\nrh{}}=125}}
    \label{fig:area_overhead}
\end{figure}

{Read disturbance mitigation techniques commonly have a trade-off between implementation complexity and execution time overheads. Figure~\ref{fig:area_overhead} shows the relationship between implementation complexity, represented by the area overhead of a mitigation technique, and its execution time overhead compared to a baseline without read disturbance mitigation.}
{We use the numbers reported in the original works for the area overheads, scaled for a two-channel system \leavevmode{to be consistent with our simulated configuration. To ensure a consistent area comparison, we normalize all hardware estimates to a 22nm technology node using DeepScaleTool~\cite{deepscaletool} where the original works report a different technology node.} For each mechanism, the figure shows two points: an implementation with priority-scheduled PROs, and \X{}.}

We observe that the \leavevmode{performance} benefits \leavevmode{of} \X{} are more pronounced for mitigation techniques that have a relatively low area overhead and, correspondingly, a high execution time overhead. Due to \X{}'s low\leavevmode{, constant} area overhead, for most mitigation techniques, we only observe a minor shift in \leavevmode{area} complexity when used alongside \X{}. \leavevmode{Compared to \leavevmode{Priority Scheduling}}, we observe a considerable reduction in execution time overhead \leavevmode{with} \X{}. We conclude that implementing especially low-complexity mitigation techniques with \X{} can make them considerably more viable for \leavevmode{read disturbance mitigation}, even at low \leavevmode{\nrh{}} values.

\subsection{{Comparison to Insecure Configuration}}
\begin{figure}[h]
    \centering
    \includegraphics[width=0.9\linewidth]{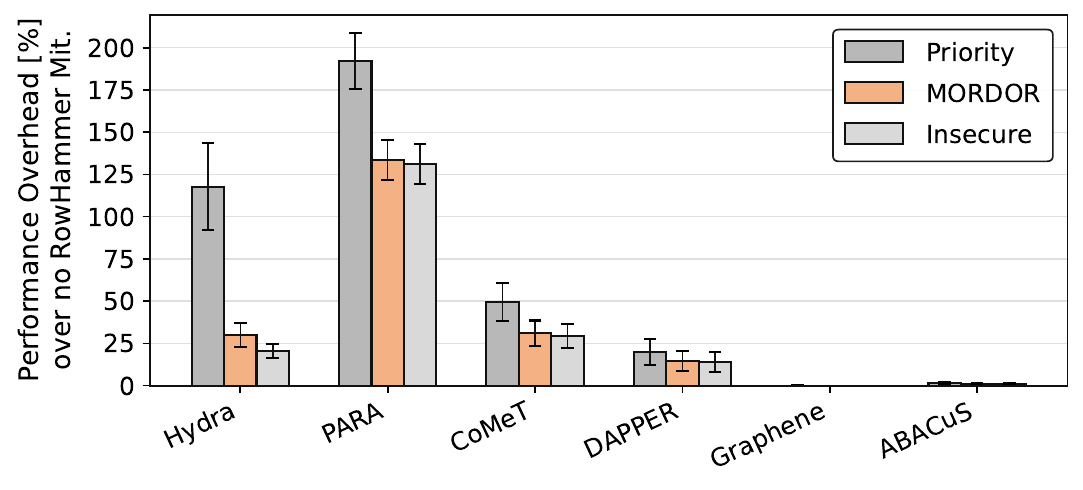}
    \caption{Performance overhead of priority scheduling, \X{}, and the insecure configuration at \leavevmode{\nrh{}}=125}
    \label{fig:insecure_comp}
\end{figure}

{We compare the performance overhead over a baseline implementation without any read disturbance mitigation for \X{} and the Insecure configuration for scheduling PROs \leavevmode{(described in §\ref{sec:mot_res})}, \textcolor{black}{to show that \X{} addresses the data integrity problems of the Insecure configuration \emph{without} sacrificing much performance}. Figure~\ref{fig:insecure_comp} shows the performance overhead of Priority Scheduling, \X{}, and Insecure over a baseline without any read disturbance mitigation for Hydra, PARA, CoMeT, DAPPER, Graphene, and ABACuS at \leavevmode{\nrh{}=125} in the 8-core setting.}

We observe that \X{} has a very low performance overhead over \textcolor{black}{Insecure}, \textcolor{black}{while, unlike \textcolor{black}{Insecure}, maintaining data integrity}. Specifically, we observe the highest overhead for Hydra, with the average performance overhead increasing by 5.2\% for \X{} \textcolor{black}{over Insecure}. For all other mitigation techniques, we observe a sub-1.2\% performance overhead of \X{} over \textcolor{black}{Insecure}. 
\leavevmode{\X{} provides data integrity through two mechanisms. 1)~blocking demand requests from activating rows with outstanding PROs and 2)~reducing the PRT by $2\cdot blast\_radius$ to account for PROs being reordered during PROQ-capacity PRO promotions (as described in §\ref{sec:PROQ_cap}). For 1), \X{} maintains the benefits of the Insecure configuration for all requests without outstanding PROs, so requests \leavevmode{\emph{not}} targeting such rows can proceed as in the Insecure configuration, without risk to data integrity. For 2), with a blast radius of two, the impact of reducing the PRT remains small for all evaluated read disturbance mitigation techniques.}
\leavevmode{We conclude that \X{} maintains most of the performance of the Insecure configuration while also maintaining the data integrity guarantees of the read disturbance mitigation technique \X{} is integrated with.}

\subsection{Sensitivity Studies}
\begin{figure}[h]
    \centering
    \includegraphics[width=0.9\linewidth]{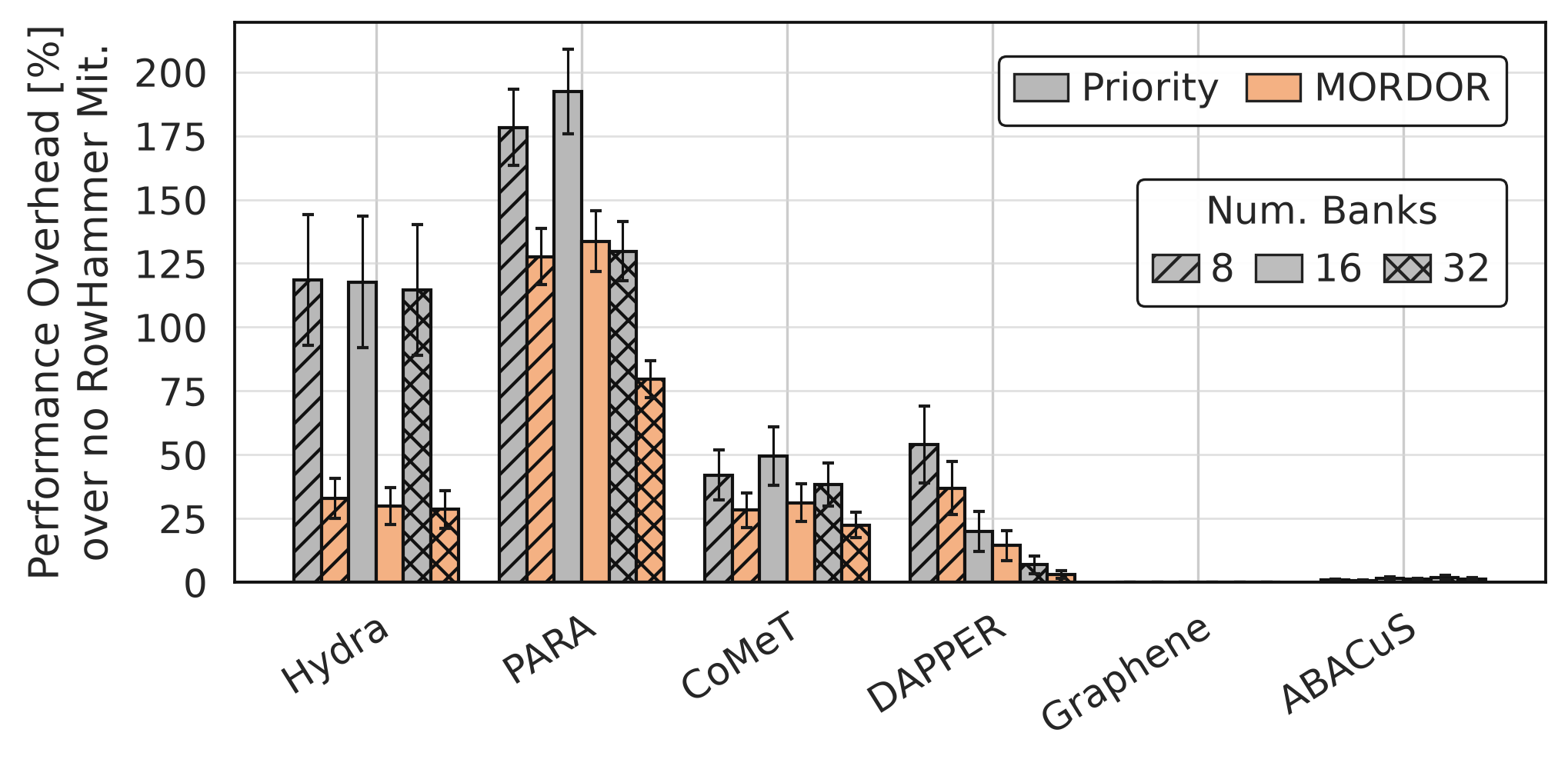}
    \caption{Performance overheads of read disturbance mitigation techniques with priority-scheduling and \X{} for 8, 16, and 32-bank configurations at \leavevmode{\nrh{}}=125}
    \label{fig:bank-study}
\end{figure}

\textbf{\leavevmode{Number of Banks.}}
\leavevmode{We evaluate two additional DRAM configurations to study \X{}'s sensitivity to the number DRAM banks}. In addition to the 16-bank configuration \leavevmode{described in §\ref{sec:meth}}, we also evaluate an 8- and 32-bank configurations. The remaining configuration parameters remain the same as given in Table~\ref{tab:system_config}. Figure~\ref{fig:bank-study} shows performance overhead of read disturbance mitigation techniques over a baseline without any read disturbance mitigation when used with Priority Scheduling and when used with \X{} at \leavevmode{\nrh{}=125} in an 8-core setting.

We observe noticeable \leavevmode{overhead reduction from \X{} over Priority Scheduling} for Hydra, PARA, CoMeT, and DAPPER in all configurations. Across all evaluated mitigation techniques and workloads, \X{} provides an average speedup of 13.7\%. 14.5\%, and 14.8\% for the 8-, 16-, and 32-bank configuration, respectively. 
\leavevmode{Different mitigation techniques exhibit different performance trends when used with \X{} as the number of banks varies. Since the number of banks affects the PRO intensity and scheduling behavior of mitigation techniques in different ways, \leavevmode{there are fewer or more optimization opportunities for \X{}}.}
\leavevmode{We conclude that \X{}'s performance benefits depend mainly on PRO-intensity, rather than the number of banks.}

\begin{figure}[h]
    \centering
    \includegraphics[width=1.0\linewidth]{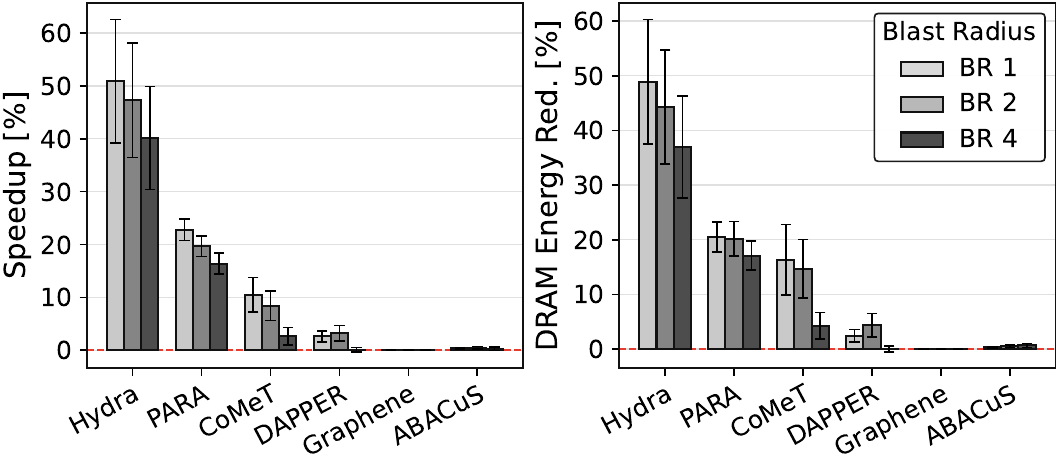}
    \caption{Speedup and DRAM energy reduction from \X{} over priority scheduling at \nrh{}=125}
    \label{fig:blast_radius}
\end{figure}

\textbf{Effect of Blast Radius.} 
We evaluate \X{} for mitigation blast radius values of 2 and 4, meaning 2 or 4 rows being refreshed on either side of the aggressor when it receives a PRO. A higher blast radius yields a higher PRO latency, as more victim rows need to be refreshed. Figure~\ref{fig:blast_radius} shows the speedup and DRAM energy reduction \X{} provides compared to \leavevmode{\leavevmode{Priority Scheduling} in an 8-core setting} at a \leavevmode{\nrh{}=125}. 
\leavevmode{As we describe in §~\ref{sec:PROQ_size}, in order to maintain data integrity with a fixed-sized PROQ, $2\cdot blast\_radius -1$ entries must be left available in the PROQ at all times. When the PROQ only has $2\cdot blast\_radius -1$ entries left, the head of the PROQ is priority scheduled. To maintain fairness between the blast radius configurations with respect to PROQ-capacity, we increase the size of the PROQ for blast radius values of 2 and 4 to 34 and 38, respectively ($30 + 2\cdot blast\_radius$). For blast radius one, the PROQ remains at 32 entries.}%

\leavevmode{We make two observations from Figure~\ref{fig:blast_radius}. First, with increasing mitigation blast radius, the speedup and DRAM energy reduction of \X{} decrease. This can be attributed to the \nrh{} reduction necessary for maintaining data integrity. Specifically, across all mitigation techniques, we observe an average cycle count reduction of 14.5\%, 13.2\%, and 9.9\% for blast radius values 1, 2, and 4, respectively. Second, for Hydra, PARA, and CoMeT, we still observe noticeable speedup and DRAM energy reduction from \X{} for all blast radius values. For DAPPER, we observe significant speedup and DRAM energy reduction for blast radius values 1 and 2, but more modest reduction at a blast radius of 4. 
\leavevmode{Out of Hydra, PARA, CoMeT and DAPPER, DAPPER is the \emph{least} PRO-intensive mitigation technique. Thus it is more affected by the blast radius increase.}}
\leavevmode{We conclude that even at higher blast radius \leavevmode{values,} \X{} can be very beneficial for PRO-intensive read disturbance mitigation techniques.}

\leavevmode{\textbf{Conclusion.} Based on our sensitivity studies, we conclude that \X{} reduces the execution time and DRAM energy overheads of PRO-intensive read disturbance mitigation techniques across a wide variety of configurations.}
\newpage

\subsection{DRFM Address-Setup Latency}
\label{sec:drfm-act-sensitivity}
\begin{figure}[h]
    \centering
    \includegraphics[width=1\linewidth]{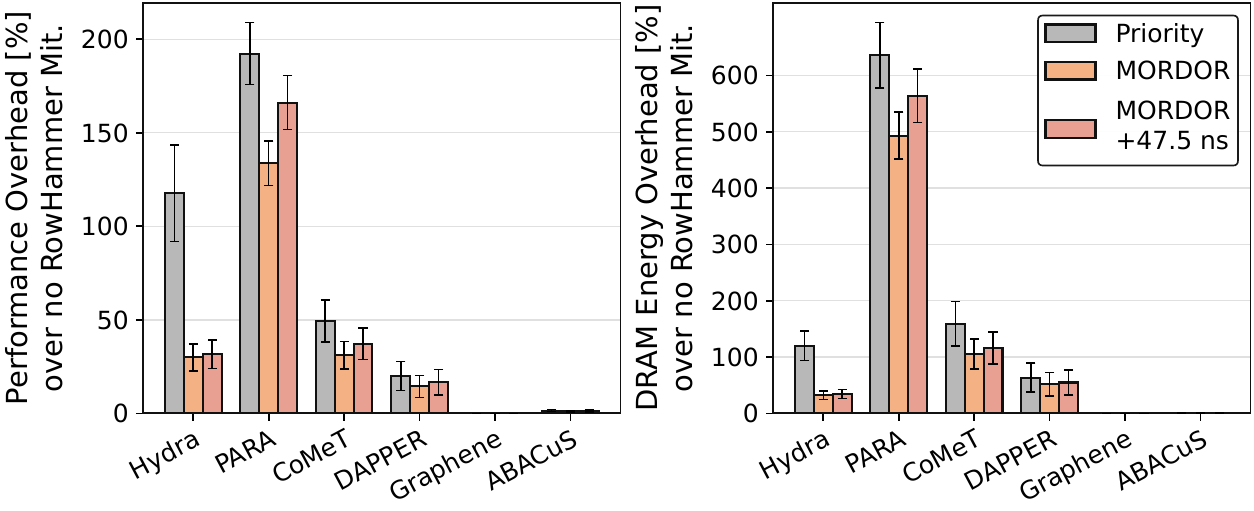}
    \caption{\leavevmode{Performance} and DRAM-energy overheads of \leavevmode{read disturbance mitigation techniques \leavevmode{with Priority Scheduling}, with MORDOR, and with MORDOR and the additional DRFM setup latency}, at \leavevmode{\nrh{}=125}}
    \label{fig:drfm-act-sensitivity}
\end{figure}

\leavevmode{When PROs are issued using the DRFM command~\cite{jedec},} \leavevmode{when using Priority Scheduling}, a DRFM is issued immediately after the selected aggressor row is closed\leavevmode{. This allows} the DRAM to retain the aggressor address in its per-bank DRFM Address Register (DAR) and identify the corresponding victim rows~\cite{jedec2020ddr5,taneja2025dream}. MORDOR decouples the detection and mitigation of an aggressor row. When the delayed DRFM is eventually issued, the \leavevmode{address of the} target row may \leavevmode{no longer} be available \leavevmode{in the DAR}. A \leavevmode{JEDEC}-compliant implementation can restore the target address through an explicit \leavevmode{activate} \leavevmode{command (ACT)} followed by \leavevmode{a precharge command (PRE)} before the DRFM.
To evaluate the overhead of issuing an \leavevmode{ACT-PRE command sequence} before a target DRFM, as described in §\ref{sec:mech}, we conservatively increase the DRFM latency for \X{} by one \leavevmode{DDR5} row-cycle interval (47.5ns)\leavevmode{~\cite{jedec, ramulator2github}}. \leavevmode{We further} decrease \leavevmode{\nrh{}} by one to account for the extra activation of the aggressor. Specifically, we increase the DRFMsb and DRFMab latencies from 240ns to 287.5ns and from 280ns to 327.5ns, respectively\leavevmode{~\cite{jedec, ramulator2github}}. Figure~\ref{fig:drfm-act-sensitivity} shows \leavevmode{performance and DRAM energy} overheads \leavevmode{over a baseline without any read disturbance mitigation} \leavevmode{when using Priority Scheduling, \X{}, and \X{} with the DRFM setup latency} at \leavevmode{\nrh{}=125} in the 8-core setting.

With the additional latency, \X{} provides an average execution-time reduction of 11.0\% and an average DRAM-energy reduction of 11.4\% over priority scheduling, retaining 75.5\% of its execution-time benefit and 77.3\% of its energy benefit. PARA is affected most because it issues DRFM commands frequently across nearly all evaluated workloads, whereas the other mechanisms issue fewer DRFMs or issue them only for a subset of workloads. We conclude that explicitly accounting for a preceding ACT and PRE sequence reduces the magnitude of \X{}'s benefits but does \leavevmode{\emph{not}} change our overall conclusion that \X{} improves the performance and reduces the energy of Priority Scheduling.

\section{Related Work}
\label{sec:related_works}
To our knowledge, \X{} is the first work \leavevmode{that introduces a high-performance Preventive Refresh Operation (PRO) scheduling technique that is based on the} insight that \leavevmode{to guarantee data integrity at high performance,} a PRO issued to an aggressor row must \leavevmode{\emph{only}} be prioritized over a demand request if the demand request activates that \leavevmode{\emph{exact}} aggressor row. 
\X{} is implemented on top of a memory-controller-based read disturbance mitigation technique to improve the performance of said technique. We already evaluated \X{} alongside six relevant
state-of-the-art RowHammer defenses~\cite{qureshi2022hydra, wang2021discreet, park2020graphene, olgun2024abacus, bostanci2024comet, dapper} in \leavevmode{§}\ref{sec:results}. Here, we describe other approaches to memory request scheduling and read disturbance mitigation.

\noindent\textbf{Memory Request Scheduling.}
\leavevmode{Various} prior works~\textcolor{black}{\cite{moscibroda2007memory,mutlu2009parallelism,kim2010atlas,ghose2013criticality,nesbit2006fair,mutlu2007stall,ebrahimi2011parallel,ausavarungnirun2012staged,usui2016dash,zhao2014firm,subramanian2014blacklisting,subramanian2016bliss,mutlu2008distributed,hur2004adaptive,mukundan2012morse,kim2010thread,subramanian2013mise,subramanian2015application,muralidhara2011reducing,lee2008prefetch,jog2016exploiting,ebrahimi2012fairness,ebrahimi2011prefetch,lee2009improving}} propose memory request schedulers to improve performance and fairness in systems in which multiple programs run concurrently. These proposals are preventive-refresh-oblivious, and propose regular memory request scheduling optimizations.
Among these works, BLISS~\textcolor{black}{\cite{subramanian2014blacklisting,subramanian2016bliss}} proposes a blacklisting method that 1)~dynamically identifies applications that are memory-intensive, and 2)~blacklist the requests from these applications to improve system fairness and performance. In contrast to MORDOR, BLISS is preventive-refresh-oblivious, and does \leavevmode{\emph{not}} take the limitations of preventive refresh scheduling into account when blacklisting requests.

\noindent\textbf{On-\leavevmode{DRAM-}Die Read Disturbance Mitigation Techniques.}
On-die read disturbance mitigation techniques aim to address \leavevmode{DRAM} read disturbance within the DRAM chip itself. Per Row Activation Counting (PRAC) \leavevmode{~\cite{canpolat2025chronus, jedec, canpolat2024understanding}} is the latest read disturbance mitigation framework that is introduced in the JEDEC DDR5 standard\leavevmode{~\cite{jedec}}.
\leavevmode{PRAC implements an activation counter for each DRAM row, and thus accurately measures the activation counts of \emph{all} rows. When a row’s activation count reaches a the PRT, the DRAM chip asserts a back-off signal which forces the memory controller to issue an PRO. The DRAM chip preventively refreshes potential victim rows upon receiving the PRO}.
Recent works \textcolor{black}{\cite{qprac, moat, canpolat2025chronus, bennett2021panopticon, vittal2025mopac, woo2025mitigations}} leverage PRAC to accurately track aggressor-row activations and mitigate the victims. \leavevmode{However, PRAC requires read-modify-write operations to update the counters for every row activation, leading to significant system performance and energy overheads\cite{canpolat2024understanding}. }%

\leavevmode{While we design \X{} as a memory-controller-based technique to aid memory-controller-based read disturbance mitigation techniques, we discussed briefly in §\ref{sec:prac} how it can be used with PRAC. We leave a full explanation of how in-DRAM read disturbance mitigation techniques and memory-controller scheduling techniques should interact and be co-designed to future work. }

DRAM manufacturers implement read disturbance mitigation techniques,
also known as Target Row Refresh (TRR)\leavevmode{~\cite{jedec, frigo2020trrespass, hassan2021utrr}} in commodity DRAM chips. The implementation of specific TRR designs is \leavevmode{\emph{not}} openly disclosed\leavevmode{~\cite{frigo2020trrespass, hassan2021utrr}}.
Prior works experimentally demonstrate that specialized adversarial access patterns can bypass these mechanisms and cause bitflips\leavevmode{~\cite{frigo2020trrespass, deridder2021smash, hassan2021utrr, jattke2022blacksmith, saroiu2022price,kogler2022half,kang2024sledgehammer,meyer2026phoenix,luo2023rowpress,jattke2024zenhammer}}.

\textbf{Memory-Controller-Based Read Disturbance Mitigation Techniques.}
\textcolor{black}{Numerous prior works\leavevmode{~\mitigatingRowHammerMCAllCitations{}} extend the memory controller with functionality to mitigate RowHammer bitflips. These prior works can be grouped into two categories, counter-based\leavevmode{~\mitigatingRowHammerMCCounterCitations{}} and probabilistic mitigation techniques\leavevmode{~\probabilisticRowHammerDefenseCitations{}}. }
\leavevmode{Counter-based mitigation techniques maintain exact or approximate counts for DRAM row activations, while probabilistic mitigation techniques avoid precisely tracking the activation counts of every row and issue mitigation based the result of a probabilistic procedure. As we show in our evaluation (§\ref{sec_results}), \X{} can improve performance and reduce energy consumption for both counter-based and probabilistic mitigation techniques,} \leavevmode{since \X{} can be implemented on top of any memory-controller-based mitigation technique that issues PROs to refresh the victims of potential aggressor rows.}

\leavevmode{\textbf{Throttling-Based Read Disturbance Mitigation Techniques.}} Some prior works\leavevmode{~\cite{yaglikci2021blockhammer, breakhammer, greenfield2016throttling, saxena2022aqua, saileshwar2022randomized, fiedler2026memory, woo2022scalable, saxena2023randomized, wi2023shadow}} propose a class of mitigation techniques that prevent read disturbance errors by limiting the activation count (throttling) of a potential aggressor row, or randomly swapping frequently accessed rows, as opposed to issuing PROs.
\leavevmode{While throttling-based mitigation techniques incur similar performance overheads to counter-based mitigation techniques, they can also cause significant delays on memory demand requests\leavevmode{\cite{yaglikci2021blockhammer, saxena2022aqua, breakhammer, woo2022scalable, saxena2023randomized}}.}

\textbf{Software-Based Read Disturbance Mitigation Techniques}~\textcolor{black}{\cite{zhang2022softtrr, van2018guardion, brasser2017can, enomoto2022efficient, konoth2018zebram, aweke2016anvil, bock2019riprh,
  irazoqui2016mascat, di2023copy, li2019detecting, loughlin2023siloz, saxena2025citadel}} forgo hardware modifications in favor of software-based read disturbance mitigation strategies. 
\leavevmode{Unfortunately, these} works, cannot monitor \leavevmode{\emph{all}} memory requests. Prior works~\textcolor{black}{\cite{gruss2016rowhammer, qiao2016new, gruss2018another, kwong2020rambleed, zhang2020pthammer, tatar2018defeating, cheng2018still, frigo2018grand, lipp2018nethammer, tatar2018throwhammer}} show that these mitigation techniques can be bypassed using well-crafted adversarial memory access patterns.

\textbf{Integrity-Checking-Based Read Disturbance Mitigation Techniques}~\textcolor{black}{\cite{juffinger2023csi, manzhosov2022revisiting, fakhrzadehgan2022safeguard, qureshi2021rethinking, saxena2023pt, kim2023how, woo2023rampart, vig2018rapid, saileshwar2018synergy}} are mitigation techniques that propose to detect bitflips and potentially correct them as opposed to preventing the occurrence of bitflips. Unfortunately, correcting all read-disturbance-induced bitflips proves to be very difficult or prohibitively expensive using integrity-checking-based mitigation techniques.
\leavevmode{Integrity-checking-based mitigation techniques} can be combined with other read disturbance mitigation techniques to improve \textcolor{black}{
overall system robustness}.

\section{Conclusion}
We propose \X{}, a new memory request scheduling technique that takes \leavevmode{Preventive Refresh Operations (PROs)} off the critical path of other memory requests while maintaining data integrity against read disturbance errors. \X{} significantly improves system performance and reduces energy consumption in the presence of PROs generated by read disturbance mitigation techniques. \leavevmode{\X{} improves performance by reducing the latency of memory demand requests.}
\X{} is compatible with memory-controller-based read-disturbance mitigation techniques that use PROs for mitigation, and requires very small modifications to the memory controller hardware. \leavevmode{\X{} is openly and freely available at \boxedurl{https://github.com/CMU-SAFARI/MORDOR}.}

\section*{Acknowledgments}
{We thank the anonymous reviewers of HPCA 2025, ISCA 2026, and MICRO 2026 for their feedback.
We thank the SAFARI Research Group members for their constructive feedback and for providing a stimulating, intellectual, and scientific environment.
We acknowledge the generous gift funding provided by our industrial partners (especially Google, Huawei, Intel, and Microsoft),
which has been instrumental in enabling the research we have been conducting on Processing-in-Memory and read disturbance in DRAM, in particular,
and memory systems in general\leavevmode{~\cite{mutlu2019processing,mutlu2017rowhammer,mutlu2019rowhammer,mutlu2022modern, mutlu2025memory,
mutlu2023fundamentally,mutlu2021intelligentdate,mutlu2019enabling,
yuksel2026memory,liu2012raidr,mutlu2023retrospective,mutlu2013memory,mutlu2014research,oliveira2021damov,gomez2021benchmarking,ghose2019processing,olgun2026modern,mutlu2024memory,mutlu2015mainmemory,ahn2023retrospective,martinez2023retrospective,oliveira2022accelerating,singh2021fpgabased,cai2017flashtbd,bostanci2026re,luo2026ramulator2_1,yuksel2025columndisturb,yuksel2025pudhammer,luo2026dejavu,luo2023rowpress,kim2014flipping,
kakolyris2026columnkeeper,bera2022hermes,kanellopoulos2026revelator}}.}

\balance
\bibliographystyle{IEEEtran}
\bibliography{refs}

\newpage
\appendix
\section{Artifact Appendix}

\subsection{Abstract}
Our artifact contains the data, source code, and scripts needed to reproduce our results. We provide: 1)~the source code of our simulation infrastructure based on Ramulator 2.0 and OpenRoad, 2)~all evaluated memory traces and all major evaluation results. We provide Python scripts to evaluate and plot the results.

\subsection{Artifact check-list (meta-information)}
\begin{table}[h]
  \centering

  \scriptsize
  \setlength{\tabcolsep}{3.5pt}
  \renewcommand{\arraystretch}{1.03}

  \begin{tabularx}{\columnwidth}{
    @{}
    >{\raggedright\arraybackslash}p{0.25\columnwidth}
    >{\raggedright\arraybackslash}X
    @{}
  }
    \toprule
    \textbf{Parameter} & \textbf{Value} \\
    \midrule

    Program &
    C++20 program\newline
    Python 3 scripts\newline
    OpenROAD hardware analysis\\
    \midrule
    
    Compilation &
    CMake $\geq 3.14$\newline
    C++20-capable compiler \\
    \midrule

    Run-time environment &
    x86-64 Linux\newline
    C++20 build toolchain\newline
    Python 3\newline
    Slurm\newline
    Git \\
    \midrule
    Hardware &
    Each simulation requests 1 CPU and 6 GB RAM, no GPU\newline OpenROAD requires $\geq 8$ GB RAM\\
    \midrule

    Execution &
    Perform simulations, aggregate results, and run plotting
    and analysis scripts. \\
    \midrule
    Metrics &
    Simulated cycle count\newline
    DRAM energy overhead \\
    \midrule
    Experiments &
    Priority, MORDOR, and insecure scheduling across six RowHammer mitigations, including PRT, bank-count, blast-radius, row-policy, and latency studies \\
    \midrule
    
    Disk space &
    20 GiB \\
    \midrule
    Preparation time &
    Approximately 30 minutes \\
    \midrule
    Experiment time &
    Approximately 3-6 days for all results depending on available infrastructure\\
    \midrule
    
    Publicly available? &
    Github:\url{https://github.com/CMU-SAFARI/MORDOR.git}\newline
    Zenodo: \url{https://doi.org/10.5281/zenodo.21539624}\newline
    Traces: \url{https://doi.org/10.5281/zenodo.21514599}
    \\
    \midrule
    Code license &
    MIT \\
    \midrule
    Automation &
    Python artifact driver, Slurm job generation and
    submission, and Python post-processing \\
    \midrule
    Archived DOI &
    10.5281/zenodo.21539624\\
    \bottomrule
  \end{tabularx}
\end{table}

\subsection{Description}
\subsubsection{How To Access} The source code and scripts are available at: GitHub: \url{https://github.com/CMU-SAFARI/MORDOR.git}, Zenodo: \url{https://doi.org/10.5281/zenodo.21539624}. The evaluated memory traces are available on Zenodo at: \url{https://doi.org/10.5281/zenodo.21514599}.
\subsubsection{Hardware Dependencies}
We recommend using a PC with 8 GiB of memory. Approximately 20 GiB of disk space is needed to store the results.

\subsubsection{Software Dependencies}
\begin{itemize}[nosep]
    \item x86-64 Linux
    \item Docker, Docker Buildx
    \item GNU Make, CMake 3.14+
    \item C++20
    \item Python 3.9+ (including venv and pip)
    \item OpenSSH client
    \item Git
    \item Bash
    \item rsync
\end{itemize}

\subsubsection{Benchmarks}
We use workload memory traces collected from
SPEC2006, SPEC2017, TPC, MediaBench, and YCSB benchmark suites. These traces are available at \url{https://doi.org/10.5281/zenodo.21514599}. Install scripts will download and extract the traces.

\subsection{Installation}
The following steps will download the repository
for the experiments. Clone and verify the artifact with:
{\small
\begin{verbatim}
$ git clone https://github.com/CMU-SAFARI/MORDOR.git
$ cd MORDOR
$ sha256sum -c CHECKSUMS.sha256
\end{verbatim}}

\subsection{Evaluation and Expected Results}
The following steps reproduce Table 1, Figure 2, and Figures 5-14 in the paper.
\subsubsection{Reproducing Table 1}
The following script builds the Docker image and runs the full experiment to reproduce Table 1 in the paper. From the cloned git repository, run:
{\small\begin{verbatim}
$ cd mordor_hw_ae
$ ./reproduce_table1.sh
\end{verbatim}}
All generated reports and the
complete console log are stored in the \cmd{out/} directory.
If Docker requires elevated privileges, use:
{\small\begin{verbatim}
$ ./reproduce_table1.sh --sudo
\end{verbatim}}
The experiment may take several hours,
depending on the available compute resources. 
The default invocation reproduces the configurations used to obtain the area
and power values reported in Table~1. 

\subsubsection{Reproducing Figures 2 and 5-14}
\emph{We strongly recommend using Slurm for running experiments in bulk.}

\textbf{Reproduction using provided infrastructure access.}
Evaluators may reproduce the data-derived Figures 2 and 5–14 using our prepared
Slurm infrastructure. The required SSH private key is provided separately
through HotCRP and is not included in the public artifact.
After cloning the artifact, the evaluator installs the key as \cmd{credentials/ae\_cluster\_key} and sets up the remote environment from the \cmd{MORDOR} directory using: 
{\small\begin{verbatim}
$ install -m 600 /path/to/provided-private-key 
    credentials/ae_cluster_key
$ ./setup_ae.sh
\end{verbatim}}
Setup synchronizes the local GitHub checkout to the cluster, excluding the SSH key and every generated output. It downloads the trace archive from Zenodo, verifies its SHA-256 checksum, and extracts the 55 paper traces.

To start the experiments to reproduce \emph{all} results in the paper, run:
{\small\begin{verbatim}
$ ./reproduce.py cluster submit
\end{verbatim}}
Depending on the available compute resources, we expect the experiments to complete in 3-5 days. To check on trace completeness, use: 
{\small\begin{verbatim}
$ ./reproduce.py cluster progress
\end{verbatim}}
Once progress reports that every selected result is valid, synchronize results and generate the data-derived paper figures using: 
{\small\begin{verbatim}
$ ./reproduce.py cluster fetch
$ .venv/bin/python reproduce.py figures
\end{verbatim}}

\cmd{cluster fetch} is completion-gated. It refuses to present an incomplete experiment as a final reproduction. \cmd{--allow-incomplete} is available for diagnosis.

\textbf{Generic Slurm cluster reproduction.} 
Clone the repository on a shared filesystem visible from the Slurm compute nodes. Copy and edit the generic profile:
{\small\begin{verbatim}
$ cp artifact_evaluation/
    generic_slurm_config.yaml \
   artifact_evaluation/
    execution_config.yaml
\end{verbatim}}

Download and verify the 55 traces, check for Slurm/build dependencies, and build Ramulator:
{\small\begin{verbatim}
$ .venv/bin/python reproduce.py slurm setup \
  --profile artifact_evaluation/
    execution_config.yaml
\end{verbatim}
}
Plan a small cohort for traces to evaluate before submitting it:
{\small\begin{verbatim}
$ .venv/bin/python reproduce.py slurm plan \
  --profile artifact_evaluation/
    execution_config.yaml \
  --classes main --traces 429.mcf
$ .venv/bin/python reproduce.py slurm submit \
  --profile artifact_evaluation/
    execution_config.yaml \
  --classes main --traces 429.mcf
\end{verbatim}
}
To run the complete paper matrix across all 55 traces on this Slurm cluster,
omit both \cmd{--classes} and \cmd{--traces}.

\textbf{Local reproduction.}
The local backend needs no Slurm installation.
Prepare the traces and simulator, then plan and run a small cohort:
{\small\begin{verbatim}
$ .venv/bin/python reproduce.py local setup \
  --profile artifact_evaluation/
    execution_config.yaml
$ .venv/bin/python reproduce.py local plan \
  --profile artifact_evaluation/
    execution_config.yaml \
  --classes main --traces 429.mcf
$ .venv/bin/python reproduce.py local run \
  --profile artifact_evaluation/
    execution_config.yaml \
  --classes main --traces 429.mcf
\end{verbatim}}
To run the complete paper matrix across all 55 traces locally, omit both
\cmd{--classes} and \cmd{--traces}.

\subsection{Methodology}
Submission, reviewing and badging methodology:

\begin{itemize}
  \item \url{https://www.acm.org/publications/policies/artifact-review-and-badging-current}
  \item \url{https://cTuning.org/ae}
\end{itemize}

\end{document}